\documentclass[preprint,3p,times, twocolumn]{elsarticle}

\usepackage{amsmath,amsfonts}
\usepackage{algorithmic}
\usepackage{algorithm}
\usepackage{array}
\usepackage{textcomp}
\usepackage{stfloats}
\usepackage{url}
\usepackage{verbatim}
\usepackage{graphicx}
\usepackage{subfig}
\usepackage{float}
\usepackage{textgreek}
\usepackage{multirow}
\usepackage{rotating}
\usepackage{tabularx}

\journal{be reviewed}

\begin{document}

\begin{frontmatter}




\title{Distilled Mid-Fusion Transformer Networks for Multi-Modal Human Activity Recognition}

\author[label1]{Jingcheng Li\corref{cor1}}
\ead{jingcheng.li@unsw.edu.au}
\author[label1,label2]{Lina Yao\corref{cor1}}
\ead{lina.yao@unsw.edu.au}
\author[label1]{Binghao Li}
\author[label1]{Claude Sammut}


\address[label1]{University of New South Wales, Sydney, 2052, NSW, Australia}
            
\address[label2]{CSIRO's Dats 61, Sydney, 2015, NSW, Australia}

\cortext[cor1]{Corresponding author}


\begin{abstract}
Human Activity Recognition is an important task in many human-computer collaborative scenarios, whilst having various practical applications. Although uni-modal approaches have been extensively studied, they suffer from data quality and require modality-specific feature engineering, thus not being robust and effective enough for real-world deployment. By utilizing various sensors, Multi-modal Human Activity Recognition could utilize the complementary information to build models that can generalize well. While deep learning methods have shown promising results, their potential in extracting salient multi-modal spatial-temporal features and better fusing complementary information has not been fully explored. Also, reducing the complexity of the multi-modal approach for edge deployment is another problem yet to resolve. To resolve the issues, a knowledge distillation-based Multi-modal Mid-Fusion approach, DMFT, is proposed to conduct informative feature extraction and fusion to resolve the Multi-modal Human Activity Recognition task efficiently. DMFT first encodes the multi-modal input data into a unified representation. Then the DMFT teacher model applies an attentive multi-modal spatial-temporal transformer module that extracts the salient spatial-temporal features. A temporal mid-fusion module is also proposed to further fuse the temporal features. Then the knowledge distillation method is applied to transfer the learned representation from the teacher model to a simpler DMFT student model, which consists of a lite version of the multi-modal spatial-temporal transformer module, to produce the results. Evaluation of DMFT was conducted on two public multi-modal human activity recognition datasets with various state-of-the-art approaches. The experimental results demonstrate that the model achieves competitive performance in terms of effectiveness, scalability, and robustness.
\end{abstract}











\end{frontmatter}


\section{Introduction}

Human Activity Recognition (HAR) is an important task in many human-computer collaborative scenarios, which delivers a variety of beneficial applications, such as health care, assisted living, elder care, and field engineering. In the multi-modal environment, the model takes diverse activity data as the input and aims to accurately predict the activity performed by a human.

\begin{figure}
\centering 
{\includegraphics[width=0.46\textwidth]{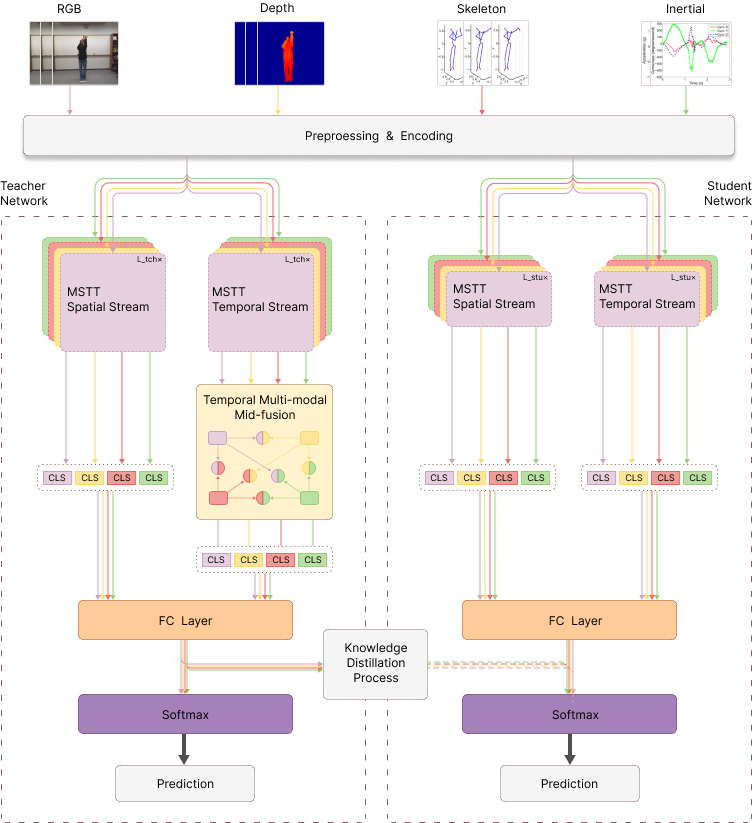}}\\
\centering
\caption{The overall framework of the DMFT model} 
\label{Fig.1} 
\end{figure}

Many existing methods have been extensively explored to resolve the Human Activity Recognition task by analyzing various uni-modal sensor data, such as RGB, depth, skeleton, inertial, and Wi-Fi data. Traditional machine learning-based approaches manually design the feature extraction methods for prediction activity labels. As those methods heavily rely on hand-craft feature engineering, they struggle to generalize well when they are applied to a new task or faced with poor data quality. Current approaches utilize deep learning techniques such as Convolutional Neural Networks (CNNs) and Long-Short Term Memory (LSTM) networks to resolve this task, which could conduct feature extraction without the need for hand-craft feature engineering. However, CNN-based models treat the temporal and spatial dimensions equally, which limits their ability to capture the sequential order of activities. Sequential models such as LSTMs focus on temporal information and ignore spatial channel information. Moreover, uni-modal approaches may not be robust enough to generalize to real-world scenarios when the input data contains noise or has low quality. For example, under low-light environments, visual-based approaches may not perform accurately as the camera could not well capture the person’s actions. Also, the inertial sensors may not be able to record and transmit the signal with negative effects of the electromagnetic environment. Thus, there is a need to explore more effective methods that can leverage both temporal and spatial information from multi-modal data sources to improve the accuracy and robustness of human activity recognition in practical settings.

As a result, multi-modal human activity recognition models have been explored to overcome the challenges of uni-modal methods. Unlike uni-modal approaches, multi-modal approaches could extract complementary information by utilizing input data from different modalities, thus producing a more robust performance.

Although multi-modal human activity recognition approaches have various advantages in extracting informative spatial and temporal features from multiple data sources and generating better results, many challenges still exist to overcome to produce more robust and accurate predictions. First, many current approaches apply a uni-modal feature extraction method for each modality stream, which not only increases model complexity but also results in extra effort to develop modality-specific methods. In the meantime, existing methods design complex network architectures to conduct feature extraction on both temporal and spatial series, which leads to an increase in the overall complexity, especially under a multi-modal scenario. If the model could not scale, it would not be suitable doe deployment in the real-world environment, as there are many low-end devices used for edge computing. Also, while multi-modal learning could benefit from utilizing comprehensive and complementary information, how to effectively conduct salient feature extraction and feature fusion still requires further exploration. Current approaches mainly conduct separate feature extraction first, then focus on late fusion among the extracted features, which may not fully share complementary information. Therefore, there is a need for further study to develop human activity recognition methods that are scalable, robust, and accurate enough to be widely deployed in real-world circumstances.

To address these challenges, we propose our Distilled Mid-Fusion Transformer (DMFT) network for multi-modal human activity recognition. This is a novel approach that resolves the Multi-modal Human Activity Recognition task in an effective, efficient, and robust way, and can be generalized into many different multi-modal environments. While achieving competitive performance in the feature extraction and fusion stage, the model is also scalable and ready to be directly deployed in real-world settings. The feature encoding layer first preprocesses the multi-modal input data, which takes the raw data of each modality as input and encodes them into a unified structure for further feature extraction. This help to improve efficiency and reduce extra data engineering effort compared with using LSTM or CNN approaches to learn feature embeddings, which can be done offline and in parallel. We then apply the Multi-modal Spatial-Temporal Transformer (MSTT) module to extract the modality-specific spatial and temporal features. By utilizing the attention mechanism, the module can extract the salient information and construct the high-level representation of the multi-modal input data. The DMFT teacher network then employs the Temporal Mid-fusion module to further extract and fuse high-level multi-modal temporal information. Unlike other approaches that use a late-fusion method, the Temporal Mid-fusion module conduct mid-fusion, which is within the feature extraction process. The DMFT student network uses a lightweight MSTT module for efficient feature extraction. Then we apply a knowledge distillation method to transfer the feature representation learned by the DMFT teacher to the smaller DMFT student model. Lastly, we use a multi-modal ensemble voting approach to aggregate the modal-specific outputs to generate the final prediction. Figure \ref{Fig.1} shows the overall DMFT framework.

We have conducted extensive experiments to evaluate the model’s performance on two public multi-modal human activity recognition datasets, UTD-MHAD \cite{chen2015utd} and MMAct \cite{kong2019mmact}, using two subject-independent evaluation protocols and various modality combinations. The results demonstrate that our proposed model has achieved competitive performance compared to the state-of-the-art approaches. Furthermore, our study suggests that by applying the knowledge distillation method we can improve the student network’s performance whilst significantly reducing the computation and space cost.

The remaining sections of this paper are organized as follows: Section 2 conducts the literature review; Section 3 presents the details of the proposed model; Section 4 introduces the datasets, the evaluation protocols and the experimental settings; Section 5 compares and reports the evaluation results and finally, Section 6 concludes the paper.

\section{Related Work}
\subsection{Human Activity Recognition}

Human Activity Recognition is a long-standing task towards human-computer interaction for years and has promising benefits in various applications. The approaches to resolving the human activity recognition task can be divided into three types: vision-based HAR, sensor-based HAR, and multi-modal HAR.

Many vision-based architectures have been extensively studied for years, which process images or videos to resolve the human activity recognition task. Traditional approaches focus on hand-craft machine learning models \cite{brand2000style,pavlovic2000learning,urtasun2008topo,wang2008gaussian,akhter2012bilinear,sutskever2008the,taylor2010dynamical}. However, those approaches often require handcraft feature engineering solutions, which are not only time-consuming but also less robust when they are deployed in a different scenario. Recently, deep learning architectures like CNN and LSTM have been widely utilized for better feature representation learning \cite{fragkiadaki2015recurrent,wang2016beyond,martinez2017human,lee2017recognition,gui2018adversarial,guo2018neural}.

Sensor-based approaches take data collected from wearable sensors, ambient sensors or object sensors as the input and conduct human activity recognition. Compared with vision-based approaches, sensor-based approaches mitigate the problems such as computational efficiency and privacy concern. Traditional machine learning approaches were also explored in the early stage \cite{fan2013human,paul2015effective,chathuramali2012faster,liu2015sensor, fallmann2016human,kabir2016two}. Recent approaches apply deep learning-based architectures, such as CNN \cite{jiang2015human,yang2015deep,peng2018aroma, chen2019multi,xue2020deepmv,bai2020adversarial} and LSTM Networks \cite{guan2017ensembles,murahari2018attention, zeng2018understanding,bock2021improving, wu2023pedal} to better extract the temporal information. 

Multi-modal approaches \cite{chen2015utd, guo2016wearable, yao2018wits, memmesheimer2020gimme,islam2020hamlet,islam2021multi,li2022multi} have been studied to resolve the human activity recognition task. Unlike uni-modal approaches, which could not generalize well due to noise or data loss, multi-modal approaches can learn robust feature representation from data of different modalities. In the meanwhile, the features of different modalities may contain complementary information thus the model can achieve improved performance. As an early attempt, Guo {\textit{et al.}} \cite{guo2016wearable} built a neural networks classifier for each modality, then used a classifier score fusion to produce the final output. Memmesheimer {\textit{et al.}} \cite{memmesheimer2020gimme} constructed signal images using the skeleton and inertial data, then treated the task as an image classification problem to predict human activities.

\subsection{The Attention Mechanism}

As an early attempt to extract the attentive information, Chen {\textit{et al.}} \cite{chen2019semisupervised} adopted a glimpse network \cite{mnih2014recurrent} to resolve sensor-based human activity recognition, where each glimpse encoded a specific area with high resolution but applied a progressively low resolution for the rest areas. 
Long {\textit{et al.}} \cite{long2018multimodal} developed a keyless attention approach to extract the spatial-temporal features from different modalities, including visual, acoustic, and segment-level features, then concatenated them to perform video classification.

Contemporary attention-based approaches \cite{chen2020metier,liu2020globalfusion,islam2020hamlet,islam2021multi,li2021two,islam2022mumu, li2022multi, suh2023tasked}, utilize self-attention methods to better extract the salient features. Islam {\textit{et al.}} \cite{islam2021multi} first built uni-modal self-attention modules to sequentially extract uni-modal spatial-temporal features, then introduced a mixture-of-experts model to extract the salient features and using a cross-modal graphical attention method to fuse the features. Their extension work \cite{islam2022mumu} added an activity group classification task and used it to guide the overall activity recognition task. Li {\textit{et al.}} \cite{li2021two} proposed a CNN augmented transformer approach to extract the salient spatial-temporal features from the channel state information (CSI) of Wi-Fi signal data to perform uni-modal human activity recognition. Their work showed the transformer’s ability to capture the salient spatial-temporal features and resolve human activity recognition tasks, but they only used one modality and Wi-Fi signals usually could contain noisy data thus the robustness would be an issue. As a result, the complexity and scalability of the model become an issue, especially under the multi-modal scenario. Whitelist self-attention-based multi-modal approaches seem to capture the complementary information and produce more robust results, they often utilize complex architectures, resulting in high space and computational complexity. As a result, these models are not suitable for real-world deployment due to the high hardware cost.

\subsection{Knowledge Distillation}

Knowledge distillation (KD) is one of the model compression methods which transfer knowledge from a computationally expensive teacher model into a smaller student network. In general, the student model is able to improve performance by learning a better feature distribution produced by a pre-trained teacher model. Currently, there are only a few KD-based methods \cite{kong2019mmact, chen2018distilling, thoker2019cross, gao2020listen,liu2021semantics,ni2022progressive,ni2022cross} that focus on multi-modal human activity recognition. Liu {\textit{et al.} \cite{liu2021semantics} first produced virtual images of inertial data, then used them to construct CNN-based teacher networks to train the student network using RGB data. Ni {\textit{et al.} \cite{ni2022progressive} developed a progressive learning method that first built a multi-teacher model using the skeleton and inertial data, then used an adaptive confidence semantic loss to let the student model adaptively select the useful information. However, those approaches only use data from 1 or 2 modalities for the teacher model and uni-modal data for the student model. As a result, the student model is still considered a uni-modal method, and the performance may suffer from noise or data loss. In this way, while they tried to conduct knowledge distillation using the multi-modal data, the models do not take advantage of multi-modal learning, which is to make use of the complementary information and improves the generalization capability. Also, some works did not follow a unified or subject-independent experimental setting in the performance comparison.

As a result, while multi-modal approaches benefit from the complementary information and may produce more robust performance, many problems still exist yet to be resolved. Firstly, existing methods often extract the salient spatial-temporal features separately for each uni-modal input data, and conduct late fusion by simply concatenating or adding the high-level features. However, the complementary information is not shared and fused in the middle stage. Secondly, while deep learning-based approaches require a huge amount of training data, it is difficult to construct well-labeled datasets in real-life, especially in a multi-modal scenario. Currently, there only exist a few multi-modal human activity recognition datasets, and the number of samples and activities is quite limited. As a result, the complex architectures may not be fully optimized and could not generalize well when more complex and new activities are introduced. Thirdly, although the SOTA multi-modal human activity recognition approaches achieve competitive performance, they introduce complex architectures which leads to high computational and space costs. In real-world scenarios, such as daily-life environments or field deployment, the devices could not afford the high cost. While knowledge distillation methods are able to reduce the model complexity, the student models still use data of a single modality, whereas the other modalities are only used to train teacher models to guide the uni-modal network. Thus there is no robust and effective approach to extracting the salient spatial-temporal multi-modal features in an efficient way.

So we propose our Distilled Mid-Fusion Transformer networks to first extract and fuse the salient spatial-temporal multi-modal features, then use a knowledge distillation method to construct a relatively simple student network to reduce the model complexity, while maintaining competitive and robust performance. To our knowledge, this is the first work that applies the knowledge distillation method to resolve the human activity recognition task in a complete multi-modal way.

\begin{figure*}
\centering 
{\includegraphics[width=0.7\textwidth]{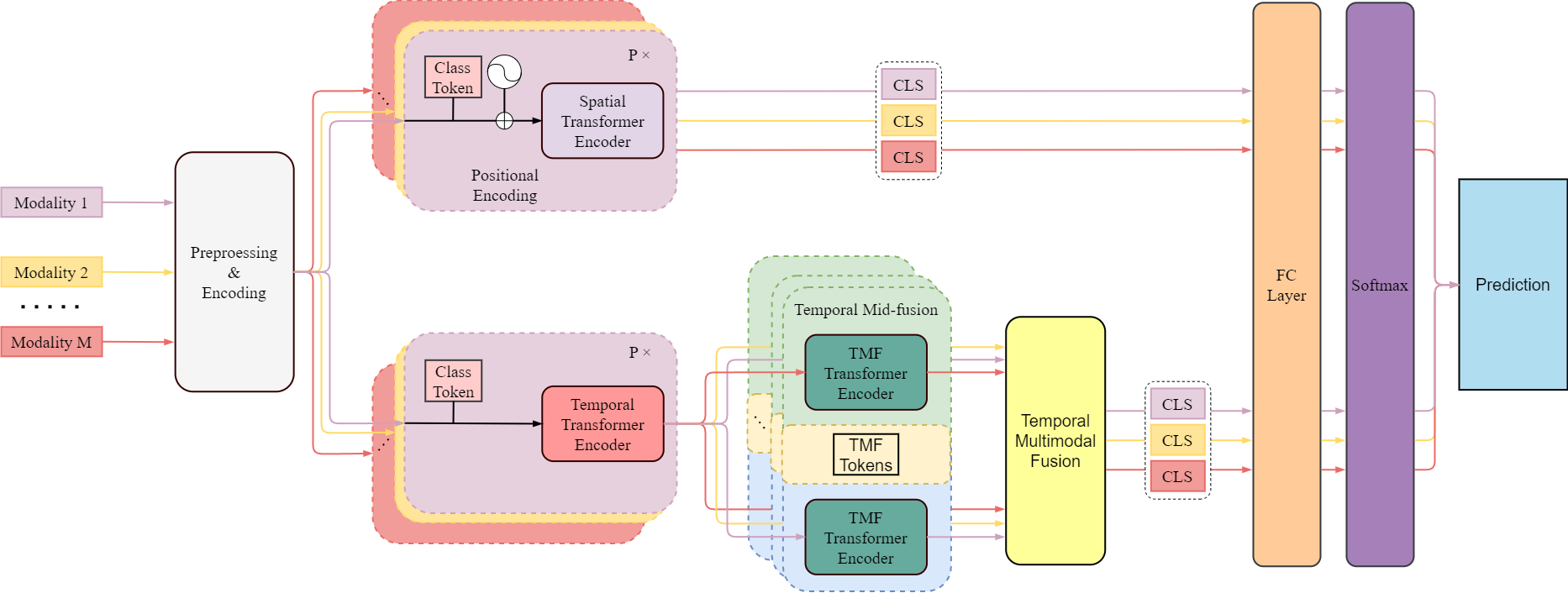}}\\
\centering
\caption{The overall framework of the DMFT teacher network} 
\label{Fig.2} 
\end{figure*}

\begin{figure*}
\centering 
{\includegraphics[width=0.7\textwidth]{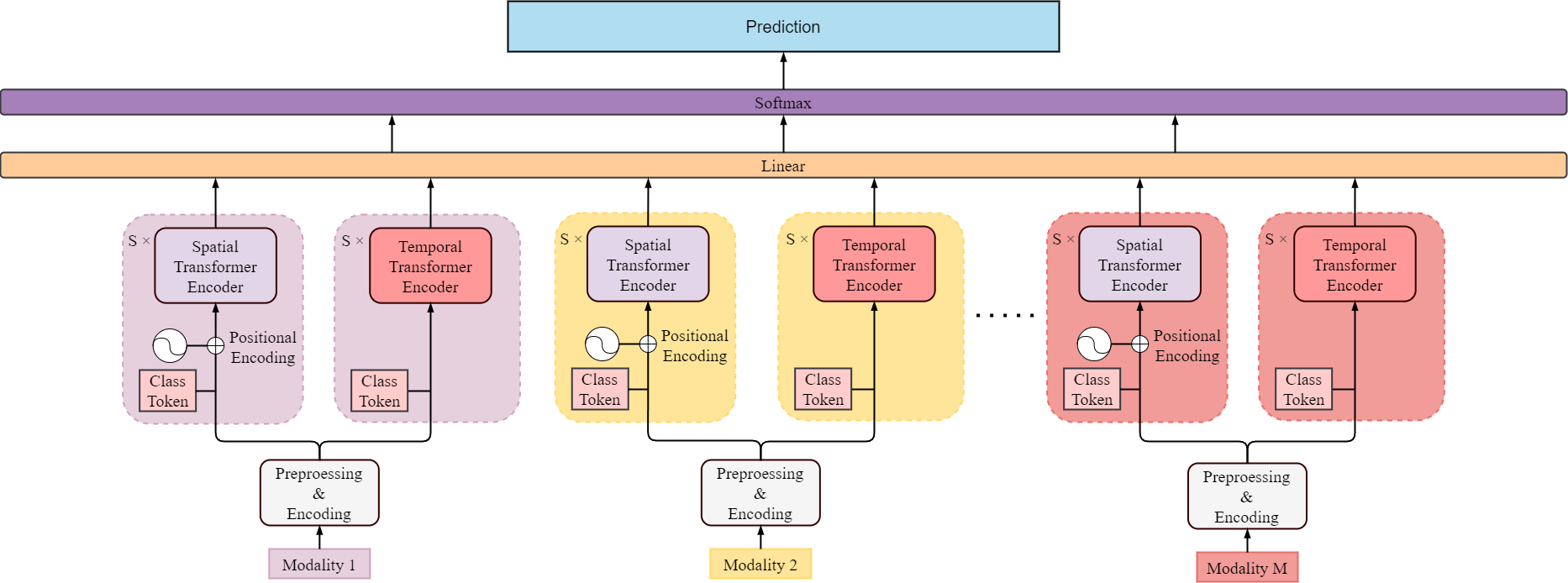}}
\centering
\caption{The overall framework of the DMFT student network} 
\label{Fig.3} 
\end{figure*}

\section{Methodology}

In this section, we propose our Multi-modal Mid-Fusion Transformer network and the knowledge distillation learning procedure for multi-modal human activity recognition. Figure \ref{Fig.2} shows the overall structure of the DMFT teacher network. Figure \ref{Fig.3} shows the overall structure of the DMFT student network. Figure \ref{Fig.1} shows the overall Knowledge Distillation process from the teacher network to the student network.

The framework contains four components:
\begin{itemize}
    \item [(i)]
    A generalized feature encoding method that receives the raw multi-modal data and encodes them into a unified structure.
    \item [(ii)]
    A multi-modal spatial-temporal transformer module and a multi-modal mid-fusion transformer encoder that serves as the teacher module, which extracts the salient spatial and temporal features, applies a temporal mid-fusion method to conduct mid-fusion among multi-modal temporal features during the feature extraction process and generate the prediction.
    \item [(iii)]
    A simple and lite multi-modal spatial-temporal transformer module that serves as the student network, generates the prediction in a scalable and efficient way.
    \item [(iv)]A knowledge distillation procedure that transfers the knowledge from a computationally expensive teacher model to a smaller student model.
\end{itemize}

First, the feature encoding layer will take the raw data of each modality as the input data and encode them into a unified structure for further feature extraction. Then we add class tokens for both the spatial stream and the temporal stream to capture the general representation of the salient features. We also add sinusoidal position encoding to the spatial stream to preserve the spatial order relationship. We apply the multi-head self-attention mechanism to better extract the salient spatial-temporal features and improve the accuracy. The DMFT teacher network utilizes the Temporal Mid-fusion Transformer module to better extract and fuse the high-level multi-modal temporal information. The DMFT student network utilizes a lite multi-modal spatial-temporal transformer module to conduct salient feature extraction in a scalable and efficient way. We then deploy a knowledge distillation procedure to transfer the feature representation learned by the DMFT teacher module to the smaller DMFT student module. Finally, we use a multi-modal ensemble voting approach to generate the overall prediction.

We elaborate on the framework in the following order: the generalized feature encoding layer, the Multi-modal Spatial-Temporal Transformer module, the Temporal Mid-fusion Transformer module, the knowledge distillation procedure from the DMFT teacher model to the DMFT student model, and the training and optimization approach.

\subsection{Generalized Encoding Layer}
Multi-modal data, such as RGB data, Depth data, Skeleton data, Inertial data, and Wi-Fi data, may have different representation structures, feature distribution, or frequencies. For example, while the Inertial data is in one dimension and has a frequency of 100HZ, the RGB data has a three-dimensional structure and is in a different frequency distribution, e.g. 24 fps. We adopt a generalized encoding layer from our previous work \cite{li2022multi}, to encode the multi-modal data into a unified representation, without applying complex modality-specific feature encoder architectures.

The reason to apply this approach is threefold. Firstly, the method does not require any handcraft feature extraction or complex feature encoders. This reduces the model complexity and makes it scalable as a new modality can be directly integrated into the network. Secondly, the method can be conducted offline in parallel and utilize pre-trained models, where the encoded features can be used by both the teacher network and the student network. This makes the approach both computationally and space efficient, which is beneficial for real-world deployment. Thirdly, our previous work \cite{li2022multi} shows that this unified encoding approach can achieve competitive results without using complex modality-specific feature encoders. Thus the approach is efficient, effective, and scalable.

For each modality $m\in M$, the input data is a set of $N$ records $R_{m} = [r_{m,1}, r_{m,2}, \dots, r_{m,N}]$. We first transfer the raw data into segments using a fixed-length sliding window and then conduct average pooling over the segment-level data. This method reduces noise and the computational and space cost \cite{li2022multi}. Moreover, to better support batch processing, we conduct a temporal alignment over the records $R_{m}$ so that each data sequence $r_{m,n}$ contains the same number of segmented patches. As the visual data (RGB and Depth) are in a two-dimensional structure over time, we utilize a pre-trained ResNet 50 model to transfer them into one-dimensional vectors over time for further feature extraction. Thus for each modality $m$, the generalized encoding layer produces the encoded input features $X_{m} = [x_{m,1}, x_{m,2}, \dots, x_{m,P_{m}}]$ of size $(B \times P_{m} \times D_{m})$, where $B$ is the batch size,$P_{m}$ denotes the length of the segmented patches, and $D_{m}$ denotes the feature dimension.

\subsection{The Multi-modal Spatial-Temporal Transformer Module}

The Multi-modal Spatial-Temporal Transformer (MSTT) module \cite{li2022multi} adapts the Transformer encoder architecture, which separately extracts the salient spatial and temporal features for each modality. While LSTM has been widely applied to resolve time series prediction tasks, the architecture suffers from the long-range dependency problem. Transformers can instead treat the input sequence as a whole and better extract the salient features by utilizing the self-attention mechanism. Moreover, unlike the traditional temporal Transformer encoder network or sequential spatial-temporal Transformer network, the results \cite{li2022multi} show that the MSTT module can better extract the salient spatial-temporal features by using separate attention modules for spatial and temporal series features.

For each modality $m$, the encoded features $X_{m}$ can be directly used as the input features $X_{m,T} = [x_{m,1}, x_{m,2}, \dots, x_{m,P_{m}}]$ for the temporal stream. For the spatial stream, a simple transpose operation can be conducted to get the input features $X_{m,S} = [x_{m,1}, x_{m,2}, \dots, x_{m,D_{m}}]$. We then add a learnable class token $x_{cls}$ to both the temporal-series features and the spatial-series features, as the class tokens can better generate the overall representation of the input features. We then add sinusoidal positional encoding to the spatial-feature stream to retain the positional information as transformer models could not capture the order information.

\begin{equation}
\begin{aligned}
PE_{(pos, 2i)}=sin(pos/10000^{2i/d_{model}})\\
PE_{(pos, 2i+1)}=cos(pos/10000^{2i/d_{model}})
\end{aligned}
\end{equation}

\begin{equation}
H_{m, S} = X_{m}^{T} = [x_{m, S, cls}, x_{m, 1}, \dots, x_{m, D_{m}}] + E_{pos}
\end{equation}
\begin{equation}
H_{m, T} = [x_{m, T, cls}, x_{m, 1}, \dots, x_{m, P_{m}}]
\end{equation}

For both the spatial stream and the temporal stream, we adopt a stack of $L_{MSTT}$-layer vanilla transformer encoders, where each layer contains a multi-head self-attention layer and a position-wise feed word layer. To extract the salient features within each Transformer encoder layer, we conduct linear projection on the input hidden embedding $H_{m}$ to create the query $Q_{m}$, key $K_{m}$ and value $V_{m}$ for each head $h$.

\begin{equation}
\begin{aligned}
Q_{m}=H_{m}W_{m}^{Q}\quad K_{m}=H_{m}W_{m}^{K}\quad V_{m}=H_{m}W_{m}^{V}\\
\end{aligned}
\end{equation}

Then we apply a scaled dot production to compute the multi-head self-attention attention scores. Where $d$ is a scaling factor to smooth the gradients. Multi-head self-attention conducts the calculation by $h$ times and then concatenates the outputs to generate the hidden embeddings for the next layer.

\begin{equation}
Attention_{m}(Q_{m}, K_{m}, V_{m})=Softmax(\frac{Q_{m}K_{m}^{\top}}{\sqrt{d_{k, m}}})V_{m}
\end{equation}

\begin{equation}
\begin{aligned}
MultiHead_{m}(Q_{m}, K_{m}, V_{m})=[head_{m, 1}, \dots, head_{m, h}]W_{m}^{O}\\
where\ head_{m, i}=Attention(Q_{m, i}, K_{m, i}, V_{m, i})
\end{aligned}
\end{equation}

The MSTT model serves as the core feature extraction module for both the teacher network and the student network to generate the salient feature representation for each modality $m$. In the next subsection, we present the Temporal Mid-fusion Transformer module, which could further conduct higher-level multi-modal fusion over the multi-modal temporal series features.

\subsection{Temporal Mid-fusion Transformer Module}

While the MSTT module could separately extract the salient multi-modal spatial and temporal features, each stream is processed separately and thus is no interaction between the multi-modal spatial-temporal features. While conducting feature addition or concatenation (known as late-fusion) is a common approach to fuse the multi-modal features, this may lose some informative information. In the meanwhile, while concatenating the multi-modal features at the beginning (known as early-fusion) could let the network take the multi-modal relationship into account, the computation cost increases quadratically, which is unrealistic for the multi-modal scenario. Also, as the multi-modal features may have different temporal lengths or feature dimensions, we cannot simply concatenate the input features and conduct feature extraction. So similar to \cite{nagrani2021attention}, we propose the Temporal Mid-fusion Transformer (TMT) Module, which further conducts multi-modal temporal feature fusion during the feature extraction step. 

For each fusion module (with a total number of the combination of the modalities), we use a set of $F$ Temporal Mid-fusion Tokens $x_{TMT}=[x_{TMT,1}, x_{TMT,2}, \dots, x_{TMT,F}]$, which serves a role similar to the class tokens, to capture the common multi-modal information whilst reducing the computational cost. For any two temporal stream features $H_{m1,T}$, $H_{m2,T}$ output by the $L_{MSTT}$-layer MSTT module, we concatenate them with the corresponding TMT tokens, then use $L_{TMT}$-layers of Transformer encoders to further extract and fuse the multi-modal features. As each temporal feature $H_{m,T}$ may have a different dimensional representation, we conduct linear projection over the TMT tokens so that they could match the features with a smaller feature dimension $D_{m}$, before and after conducting the temporal mid-fusion. This would help to conduct feature fusion among two different streams with different dimensions.

\begin{equation}
\begin{aligned}
H_{m1,T}^{TMT}=[x_{TMF,(m1,m2)},H_{m1,T}]\\
H_{m2,T}^{TMT}=[LN(x_{TMF, (m1,m2)}),H_{m2,T}]\\
where\ D_{m1} < D_{m2}
\end{aligned}
\end{equation}

Then for each fusion combination of modalities $m_{1}$ and $m_{2}$, the multi-modal Temporal Mid-fusion attention flow is structured as below.

\begin{equation}
H_{m,T}^{TMT'}=Transformer(H_{m,T}^{TMT},\theta_{m})\\
\end{equation}

\begin{equation}
x_{TMF}^{'}=Average(x_{TMF,m1}^{'},x_{TMF,m2}^{'})
\end{equation}

It is worth mentioning that when the number of modalities $M > 2$, the Temporal Mid-fusion will be done by a combination of $C_{M}^{2}$ times, which will be the number of sampling 2 combined modalities from $M$ modalities. Then for each modality $m$, we average the output of each Temporal Mid-fusion Transformer module to hierarchically fuse the multi-modal information.

\begin{equation}
H_{m,T}^{TMT,L} = Average(H_{m,T,1}^{TMT,L}, H_{m,T,2}^{TMT,L}, \dots, H_{m,T,C_{M}^{2}}^{TMT,L})
\end{equation}

\subsection{Multi-modal Knowledge Distillation}
We then construct the multi-modal teacher network, which is a combination of the MSTT module and the TMT module, to better extract and fuse the multi-modal spatial and temporal features. The overall framework of the teacher network is illustrated in Figure \ref{Fig.2}. 

For each modality $m$, the raw data are first passed through the generalized encoding layer to get the unified input features. We then apply the vanilla self-attention MSTT module with $L_{MSTT, tch}$ among the tokens $X_{m,S}$, $X_{m,T}$ to extract the salient features for both the spatial and the temporal streams. After this, we concatenate each combination of the two temporal latent features output by the MSTT module with the corresponding TMT tokens and pass them through the TMT module, where the tokens are fused and updated in accordance with the formula. 

Then for both the spatial and the temporal stream, we output the corresponding representations of the class tokens and pass them through a linear layer. For the teacher network, we apply a Softmax function among the logits and average the outputs of each modality and produce the overall prediction $Y_{t}$.

\begin{equation}
Y_{m}=Softmax(LN_{S}(h_{m, S, L, 0})) \\ 
    + \\
    Softmax(LN_{T}(h_{m, T, L, 0}^{TMT}))
\end{equation}

\begin{equation}
\hat{Y} =\sum_{m=1}^{M}Y_{m}
\end{equation}

The framework of the student network is illustrated in Figure \ref{Fig.3}. The student network is a simpler architecture that contains a $L_{MSTT, stu}$-layer MSTT module. For each modality, the input features are passed through the MSTT module and we then output the representation of the class tokens and pass them through a linear layer. Similar to the teacher network, a Softmax function is applied over the logits to generate the overall prediction.

We then apply a knowledge distillation based approach to train the student network. The training process is shown in Figure \ref{Fig.1}. We apply a Softmax operation to convert the output logits into class probabilities $P_{teacher}$, which is softened by the temperature parameter ${temp}$. Assume that for each modality m, the class probabilities output by the teacher network is denoted by $P_{teacher,S}$, $P_{teacher, T}$. The student network’s class probabilities $P_{student,S}$, $P_{student, T}$ are then optimized to match the corresponding teacher network logits distribution to predict the target class. 

\begin{equation}
P = \frac{\frac{exp(h_{i}}{temp})}{\sum_{j}exp(\frac{h_{j}}{temp})}
\end{equation}

Hinton et. al used a KullbackLeibler(KL) divergence in the loss function to conduct knowledge transfer from the teacher network to the student network so that the student network’s class probabilities will converge to those output by the teacher network.

\begin{equation}
KL(P_{student}, P_{teacher}) = \sum_{c}P_{student,c}log\frac{P_{student,c}}{P_{teacher,c}}
\end{equation}

To train the student network, we use a weighted loss which consists of the cross entropy loss $L_{CS}$ (hard loss) along with the knowledge distillation loss $L_{KD}$ (soft loss). The overall loss function $L_{student}$ to optimize the student network is defined as follows:

\begin{equation}
\begin{aligned}
L_{student}=w_{CS} L_{CS}+ \\
\sum_{m=1}^{M}w_{m,S}L_{KL}(P_{student,m,S},P_{teacher,m,S})+ \\
\sum_{m=1}^{M}w_{m,T}L_{KL}(P_{student,m,T},P_{teacher,m,T})+ \\
where\ w_{CS}+\sum_{m=1}^{M}w_{m,S}+\sum_{m=1}^{M}w_{m,T}=1
\end{aligned}
\end{equation}

Then the overall training and optimization process is to minimize the loss $L_{student}$ and generate the prediction $\hat{Y}$ for the given multi-modal input data.

\section{Experiments}

\subsection{Datasets}

We evaluate DMFT’s performance on two public benchmark datasets, UTD-MHAD and MMAct, which are the only two mainstream multi-modal human activity recognition datasets available in the area.

The UTD-MHAD \cite{chen2015utd} dataset consists of 27 activities, where each activity is performed by 8 subjects (4 males and 4 females) 4 times, resulting in 861 samples after removing the corrupted samples. The dataset includes 4 modalities, RGB, Depth, Skeleton, and Inertial. A Kinect camera is used to capture the visual information while a wearable sensor is used to record the inertial data, including acceleration, gyroscope, and magnetic data. All 4 modalities are used in our experimental setup.

The MMAct \cite{kong2019mmact} dataset consists of 35 activities, where each activity is performed by 20 subjects (10 males and 10 females) 5 times, resulting in 36K samples after removing the corrupted samples. The dataset includes 7 modalities, RGB, Skeleton, Acceleration, Gyroscope, Orientation, Wi-Fi, and Pressure. 4 cameras and a smart-glass are used to record the RGB data, while a smartphone and a smartwatch are used to record the acceleration, gyroscope, orientation, Wi-Fi and pressure data. We used acceleration, gyroscope, and orientation as the Inertial data and RGB data to conduct the experiments.

\subsection{Evaluation Protocol}

For the UTD-MHAD dataset, we use two types of cross-validation methods. First, same as the original paper \cite{chen2015utd}, we apply a 50-50 evaluation method, where the odd-numbered subjects (1, 3, 5, 7) are used for training and the even-numbered subjects (2, 4, 6, 8) are used for testing. Meanwhile, we apply a leave-one-subject-out (LOSO) protocol, where we iteratively select each subject for testing and use the other 7 subjects for training. We use Top-1 accuracy as the evaluation metric for the UTD-MHAD dataset and take the average of the results to compare the model’s performance with the other approaches.

For the MMAct datasets, we followed two cross-validation protocols proposed by the original paper \cite{kong2019mmact}. First, we use a cross-subject setting, where the first 80\% of the numbered subjects (1-16) are used for training and the rest of the numbered subjects (17-20) are used for testing. A cross-session setting is also used, where the first 80\% of the sessions for each subject are used for training and the rest sessions are used for testing. We use F1-score as the evaluation metrics for the MMAct dataset and take the average of the results to compare the model’s performance with the other approaches.

It is worth noting that the 50-50 subject setting, the LOSO setting, and the cross-subject setting are subject-independent. In the real-world scenario, the model is used to analyze the activities performed by new subjects. Subject-dependent evaluation protocols, where both the training set and the test set both contain the common information of the same subject, neglect the participant bias and may lead to a different conclusion. So we mainly apply the subject-independent setting to take the real-world variation for different subjects to better evaluate the model’s performance. Meanwhile, although the cross-session setting is subject-dependent, we include it to further demonstrate the model’s performance details.

\subsection{Experimental Settings}

Preprocessing. As some modalities may contain different types of streams, we treat them as a single modality and conduct preprocessing. For each modality, we first separate the data stream into segments using a sliding window to downsample the frequency. Then for both the RGB and Depth data, we encode them using a pre-trained ResNet50 network. For the Skeleton and Inertial data, we directly pass them through the feature extraction module. For each spatial and temporal stream, we concatenate a class token with the input data. 

We implement the model using the PyTorch framework and use Adam optimization. For the experiments on the UTD-MHAD dataset, we use an NVIDIA RTX 3090 GPU, while for the experiments on the MMAct dataset, we use an NVIDIA A40 GPU. We use Top-1 accuracy as the evaluation metrics for experiments on the UTD-MHAD dataset and F1-score as the evaluation metrics for experiments on the MMAct dataset.

\section{Results and Comparisons}

\subsection{Overall Comparison}

We evaluated the performance of DMFT by conducting experiments on two multi-modal HAR datasets: UTD-MHAD and MMAct. 

For the UTD-MHAD dataset, as mentioned above, we apply both the 50-50 subject evaluation protocol and the LOSO evaluation protocol. The experimental results on the UTD-MHAD dataset can be found in Table \ref{tab:table_1} and Table \ref{tab:table_2}. We compare our approach to the baseline approach as well as more recent multi-modal approaches. Under the 50-50 subject protocol, the results show that the DMFT teacher model outperforms the other multi-modal approaches by achieving 93.97\% accuracy with Skeleton and Inertial data. While the DMFT student model achieves 92.12\% accuracy, which is only 0.6\% lower than the predecessor MATN model. For the LOSO setting, the DMFT teacher model outperforms the other multi-modal approaches by achieving 98.20\% using RGB, Skeleton, and Inertial data. In the meanwhile, both the DMFT teacher model and the DMFT student model achieve competitive results compared to the state-of-the-art approaches.

\begin{table}[]
    \centering
        \caption{50-50 subject performance comparison on the UTD-MHAD dataset. S: Skeleton, D: Depth, I: Inertial, aug.:augumentation.}
    \label{tab:table_1}
    \resizebox{1\columnwidth}{!}{

    \begin{tabular}{ccc}
        \hline
        {\bf Method}& {\bf Modality Combination} & {\bf Accuracy (\%)} \\ 
        \hline
        MHAD \cite{chen2015utd} & I+D & 79.10 \\
        \hline
        MHAD \cite{chen2015utd} & I+D & 81.86 \\
        \hline
        Gimme Signals \cite{memmesheimer2020gimme} & I+S & 76.13 \\ 
        \hline
        Gimme Signals \cite{memmesheimer2020gimme} & I+S (data aug.) & 86.53 \\
        
        \hline
        MATN & I+S & 92.72 \\
        \hline
        DMFT (Teacher) & I+S & {\bf93.97} \\
        \hline
        DMFT (KD) & I+S & 92.12 \\
        \hline
    \end{tabular}
    }
\end{table}

\begin{table}[]
    \centering
        \caption{LOSO performance comparison on the UTD-MHAD dataset. R: RGB, S: Skeleton, D: Depth, I: Inertial.}
    \label{tab:table_2}
   
    \begin{tabular}{cccc}
        \hline
        \multirow{2}*{\bf{Method}} & \multicolumn{3}{c}{\bf{Accuracy (\%)}}\\
        \cline{2-4}
       & R+S & R+S+I & R+S+D+I\\
        \hline
        Keyless \cite{long2018multimodal} & 90.20 & 92.67 & 83.87\\
        \hline
        HAMLET \cite{islam2020hamlet}  & 95.12 & 91.16 & 90.09\\
        \hline
        Multi-GAT \cite{islam2021multi}  & {\bf96.27} & 96.75 & 97.56\\
        \hline
        Mumu   & 96.10 & 97.44 & {\bf97.60}\\
        \hline
        MATN   & 90.37 & 97.62 & 97.46\\
        \hline
        DMFT (Teacher)   & 93.06 & {\bf98.20} & 97.52\\
        \hline
        DMFT (KD)  & 90.26 & 96.52 & 96.53\\
        \hline
        
    \end{tabular}
    
\end{table}
\begin{table}[]
    \centering
        \caption{Cross-subject performance comparison on the MMAct dataset. R: RGB, I: Inertial.}
    \label{tab:table_3}
    \resizebox{1\columnwidth}{!}{

    \begin{tabular}{ccc}
        \hline
        {\bf Method}& {\bf Modality Combination} & {\bf F1-Score (\%)} \\ 
        \hline
        SMD \cite{hinton2015distilling} & I+R & 63.89 \\
        \hline
        Student \cite{kong2019mmact} & R & 64.44 \\
        \hline
        Multi-teachers \cite{kong2019mmact} & I &  62.67 \\
        \hline
        MMD \cite{kong2019mmact} & I+R &  64.33 \\
        \hline
        MMAD \cite{kong2019mmact} & I+R &  66.45 \\
        \hline
        REPDIB+MM (HAMLET) \cite{islam2022representation} & I+R &  57.47 \\
        \hline
        REPDIB+MM (Keyless) \cite{islam2022representation} & I+R &  63.22 \\
        \hline
        REPDIB+MM (REPDIB+Uni) \cite{islam2022representation} & I+R &  69.39 \\
        \hline
        HAMLET \cite{islam2020hamlet} & I+R &  69.35 \\
        \hline
        PSKD \cite{ni2022progressive} & I+R & 71.42 \\
        \hline
        Keyless \cite{long2018multimodal} & I+R & 71.83 \\
        \hline
        Multi-GAT \cite{islam2021multi} & I+R &  75.24 \\
        \hline
        SAKDN \cite{liu2021semantics} & I+R &  77.23 \\
        
        \hline
        Mumu & I+R & 76.28 \\
        \hline
        MATN & I+R & {\bf83.67} \\
        \hline
        DMFT (Teacher) & I+R & 83.29 \\
        \hline
        DMFT (KD) & I+R & 82.54 \\
        \hline
    \end{tabular}
    }
\end{table}

\begin{table}[]
    \centering
        \caption{Cross-session performance comparison on the MMAct dataset. R: RGB, I: Inertial.}
    \label{tab:table_4}
    \resizebox{1\columnwidth}{!}{

    \begin{tabular}{ccc}
        \hline
        {\bf Method}& {\bf Modality Combination} & {\bf F1-Score (\%)} \\ 
        \hline
        MMAD \cite{kong2019mmact} & I+R + RGB & 74.58 \\
        \hline
        MMAD(Fusion) \cite{kong2019mmact} & I+R + RGB & 78.82 \\
        \hline
        Keyless \cite{long2018multimodal} & I+R + RGB & 81.11 \\
        \hline
        SAKDN \cite{liu2021semantics} & I+R + RGB &  82.77 \\
        \hline
        HAMLET \cite{islam2020hamlet} & I+R + RGB & 83.89 \\
        \hline
        Multi-GAT \cite{islam2021multi} & I+R + RGB & 91.48 \\
        \hline
        Mumu & I+R + RGB & 87.50 \\
        \hline
        MATN  & I+R + RGB & {\bf91.85}\\
        \hline
        DMFT (Teacher) & I+R + RGB & 91.62 \\
        \hline
        DMFT (KD) & I+R + RGB & 91.09 \\
        
        \hline
    \end{tabular}
    }
\end{table}

\begin{figure*}
    
  \centering
  \subfloat[50-50 Teacher]{\includegraphics[width=0.23\textwidth]{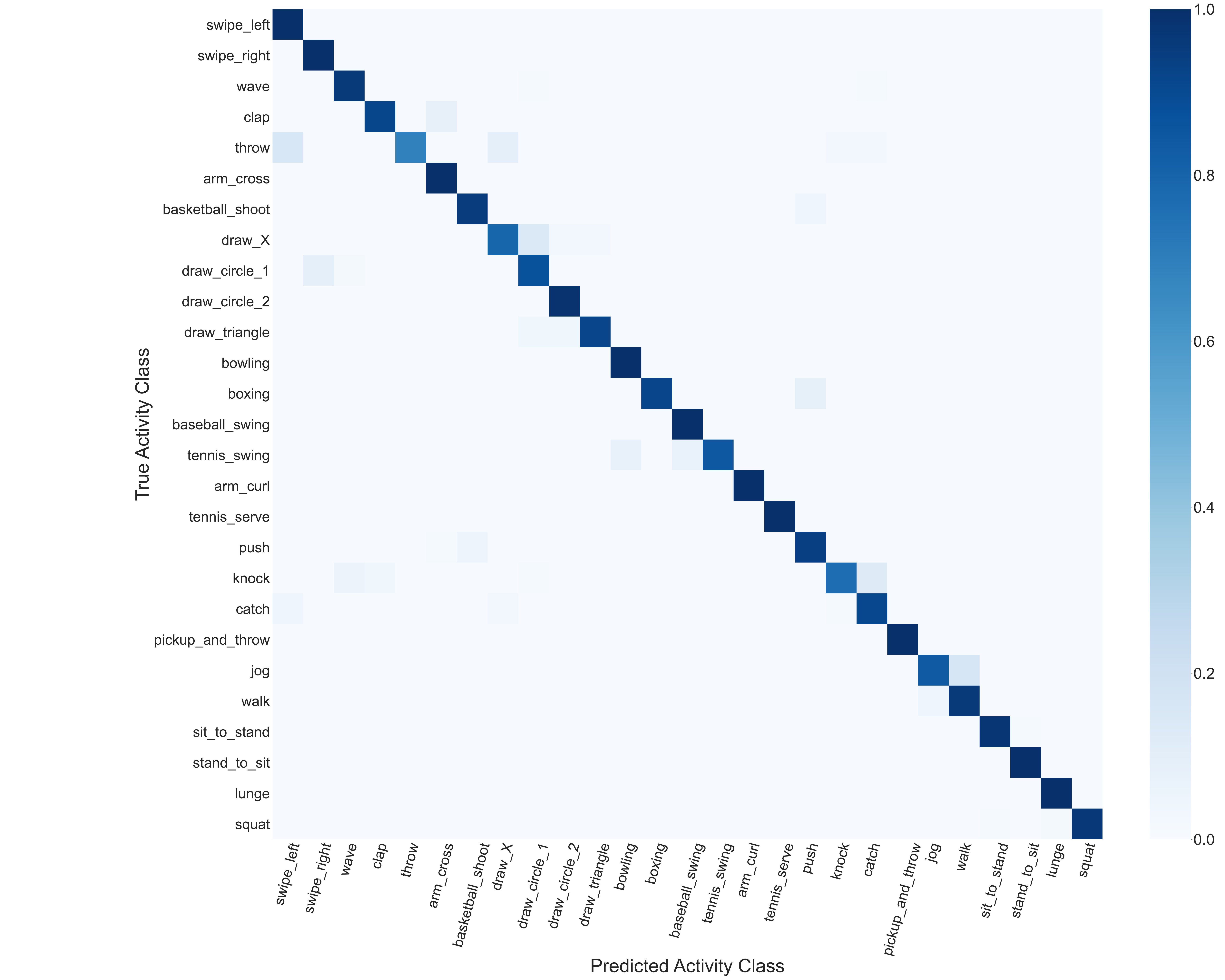}}
  \hspace{0.01\textwidth}
  \subfloat[50-50 KD]{\includegraphics[width=0.23\textwidth]{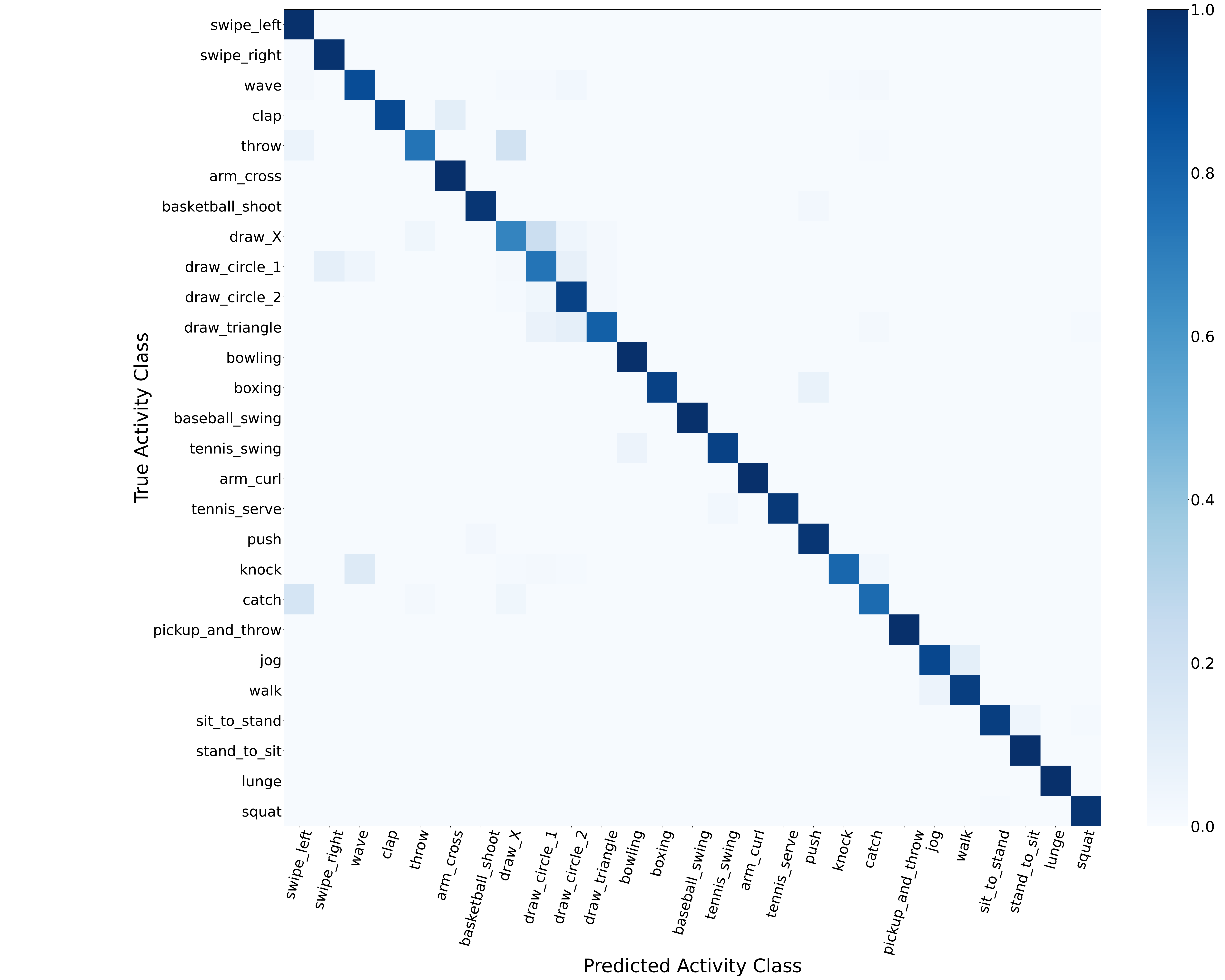}}
  \hspace{0.01\textwidth}
  \subfloat[RSDI Teacher]{\includegraphics[width=0.23\textwidth]{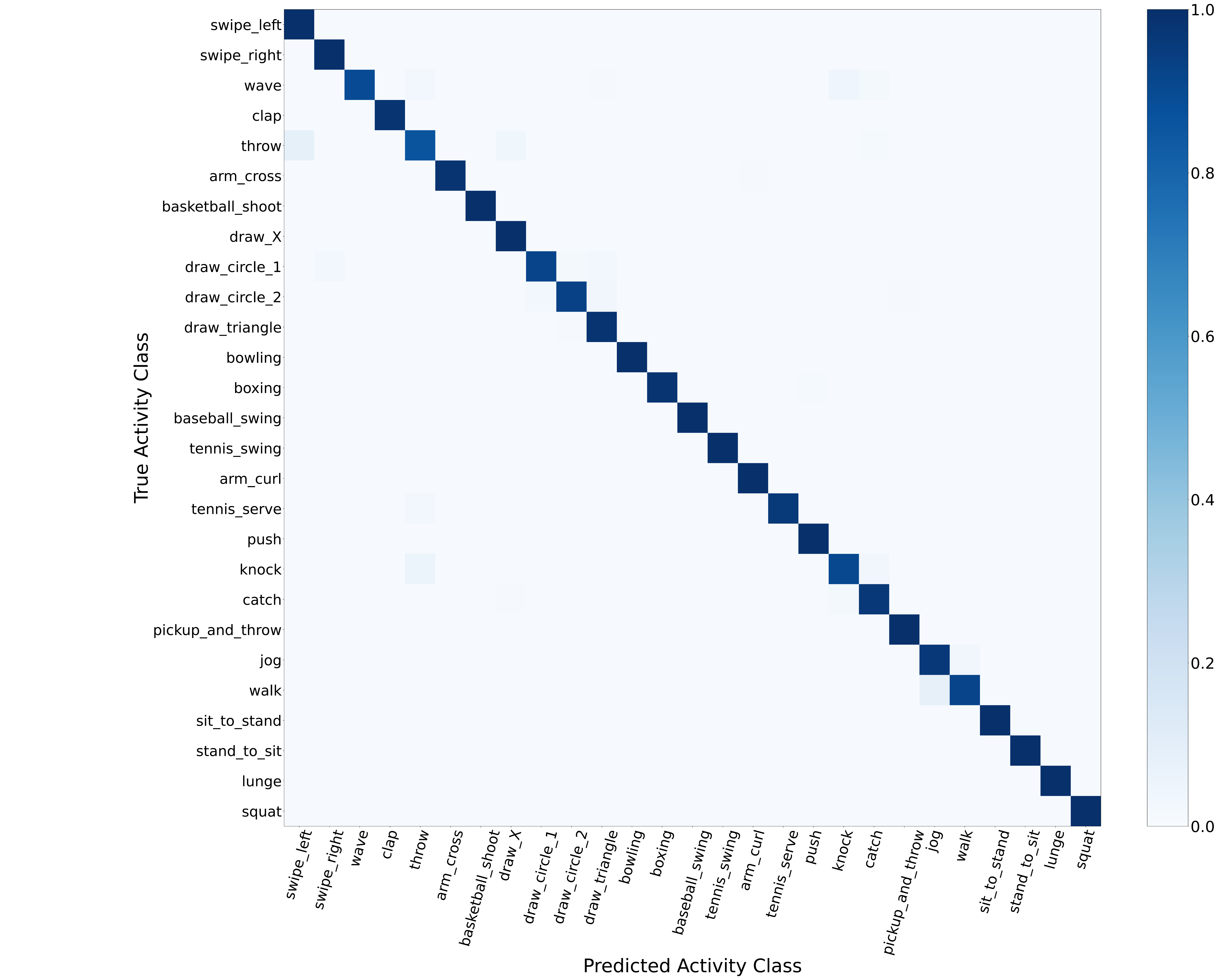}}
  \hspace{0.01\textwidth}
  \subfloat[RSDI KD]{\includegraphics[width=0.23\textwidth]{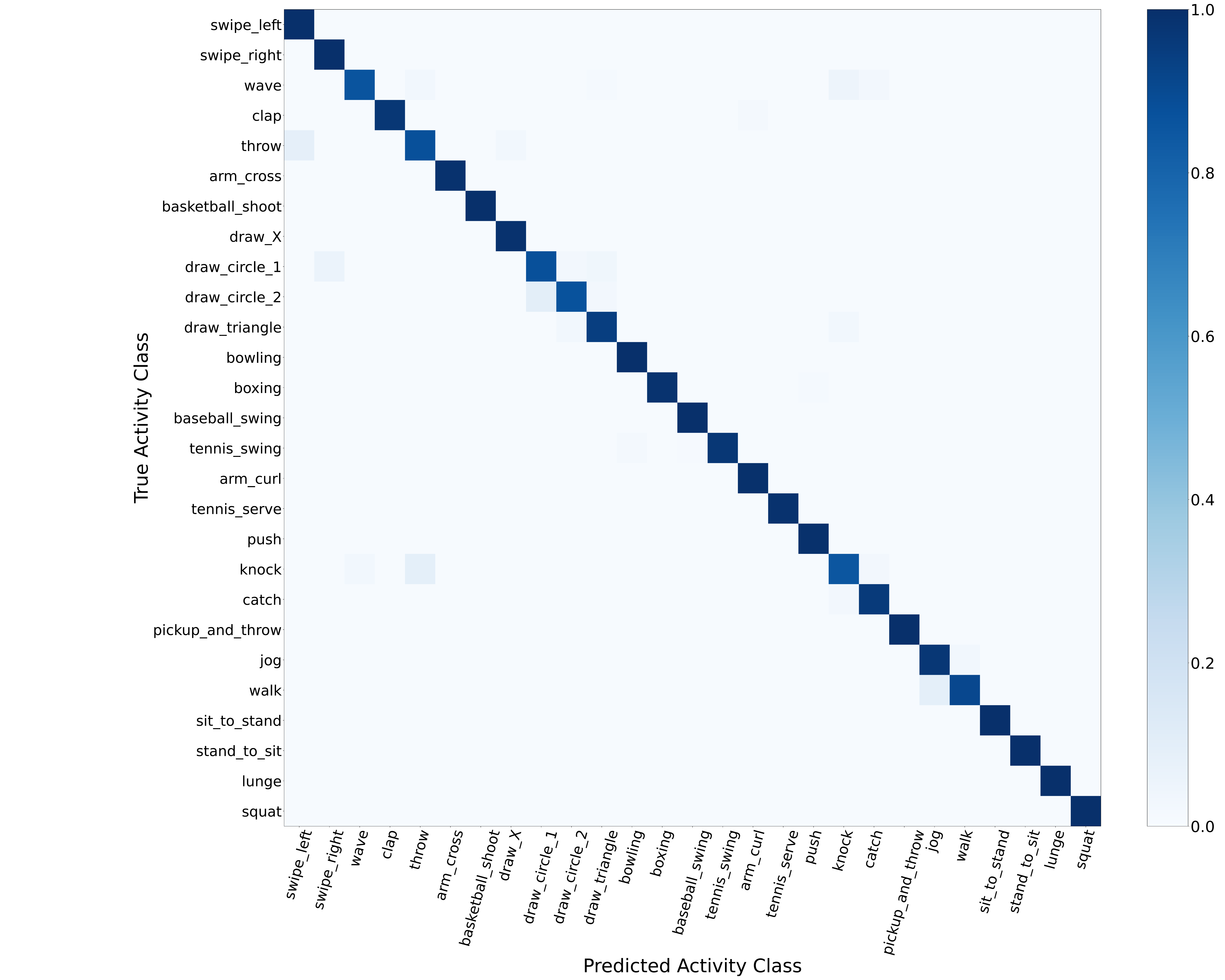}} \\

  \subfloat[session Teacher]{\includegraphics[width=0.23\textwidth]{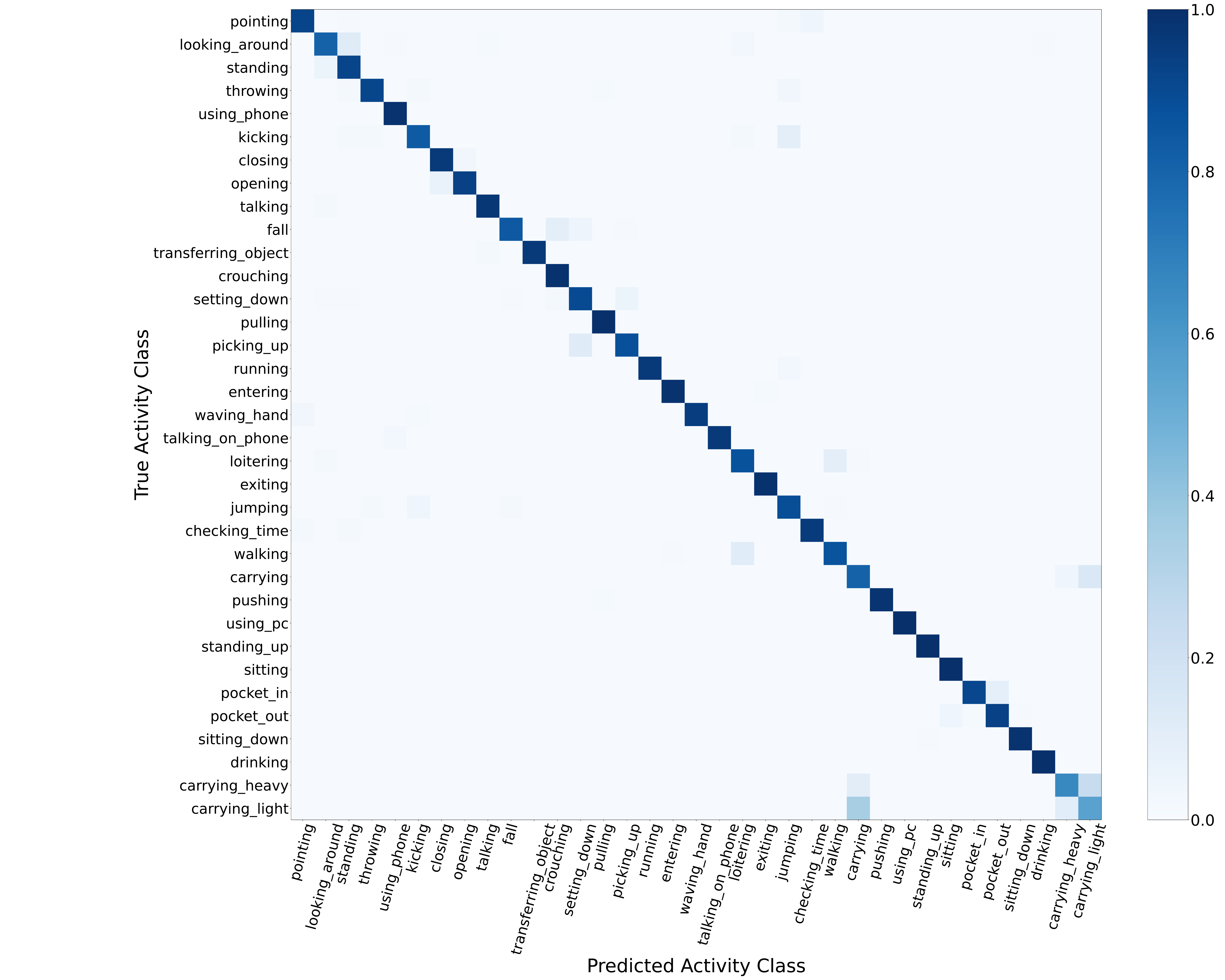}}
  \hspace{0.01\textwidth}
  \subfloat[session KD]{\includegraphics[width=0.23\textwidth]{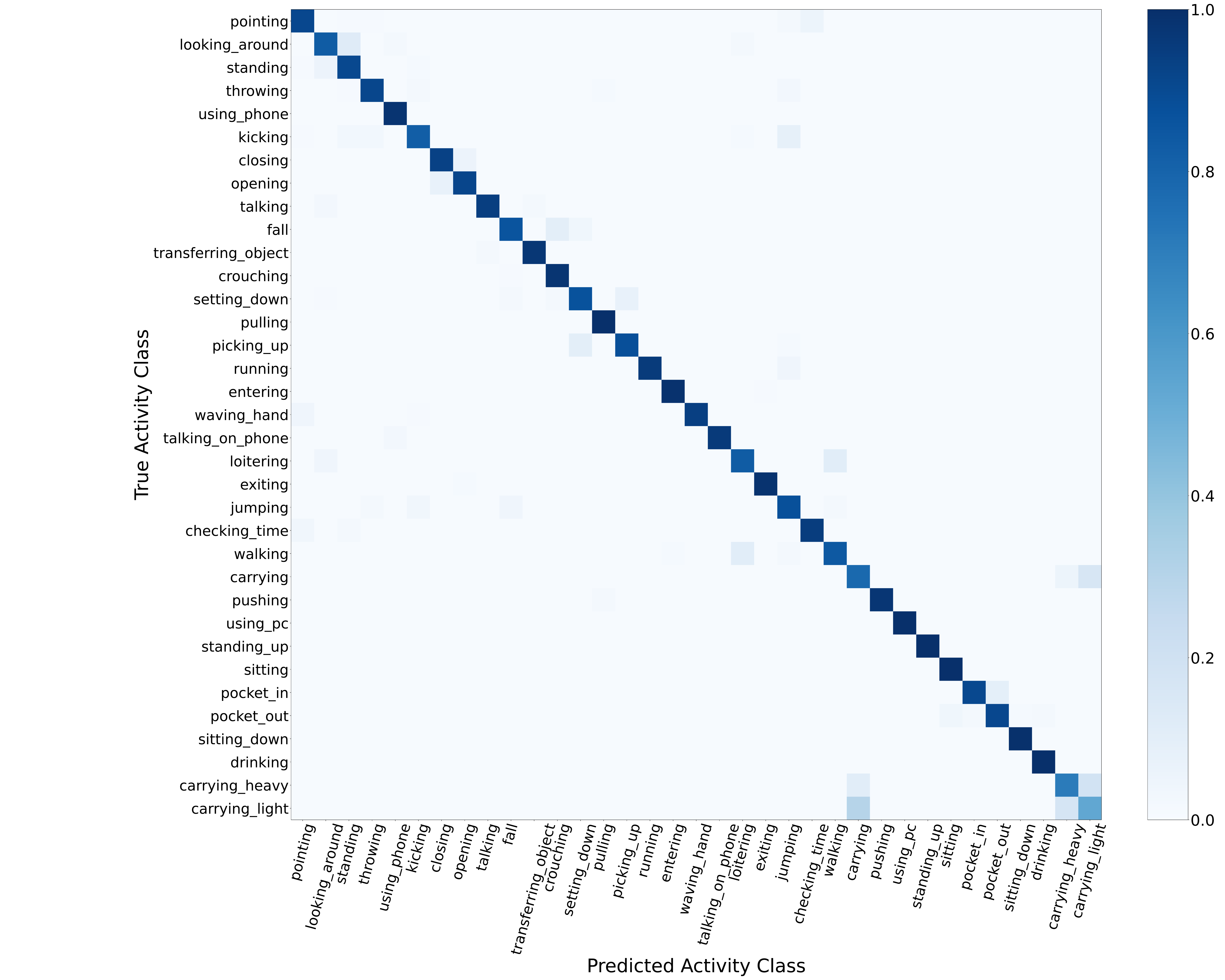}}
  \hspace{0.01\textwidth}
  \subfloat[subject Teacher]{\includegraphics[width=0.23\textwidth]{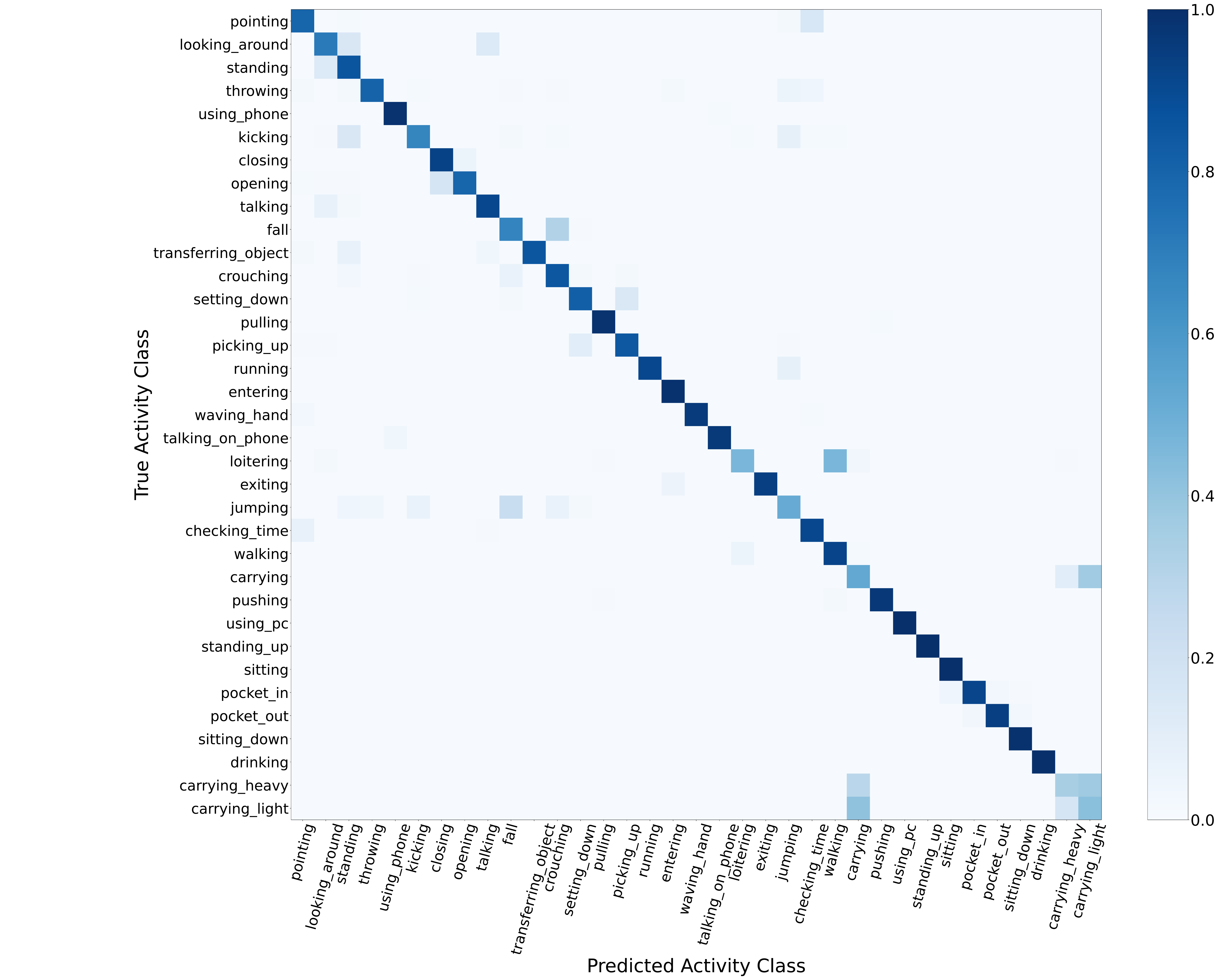}}
  \hspace{0.01\textwidth}
  \subfloat[subject KD]{\includegraphics[width=0.23\textwidth]{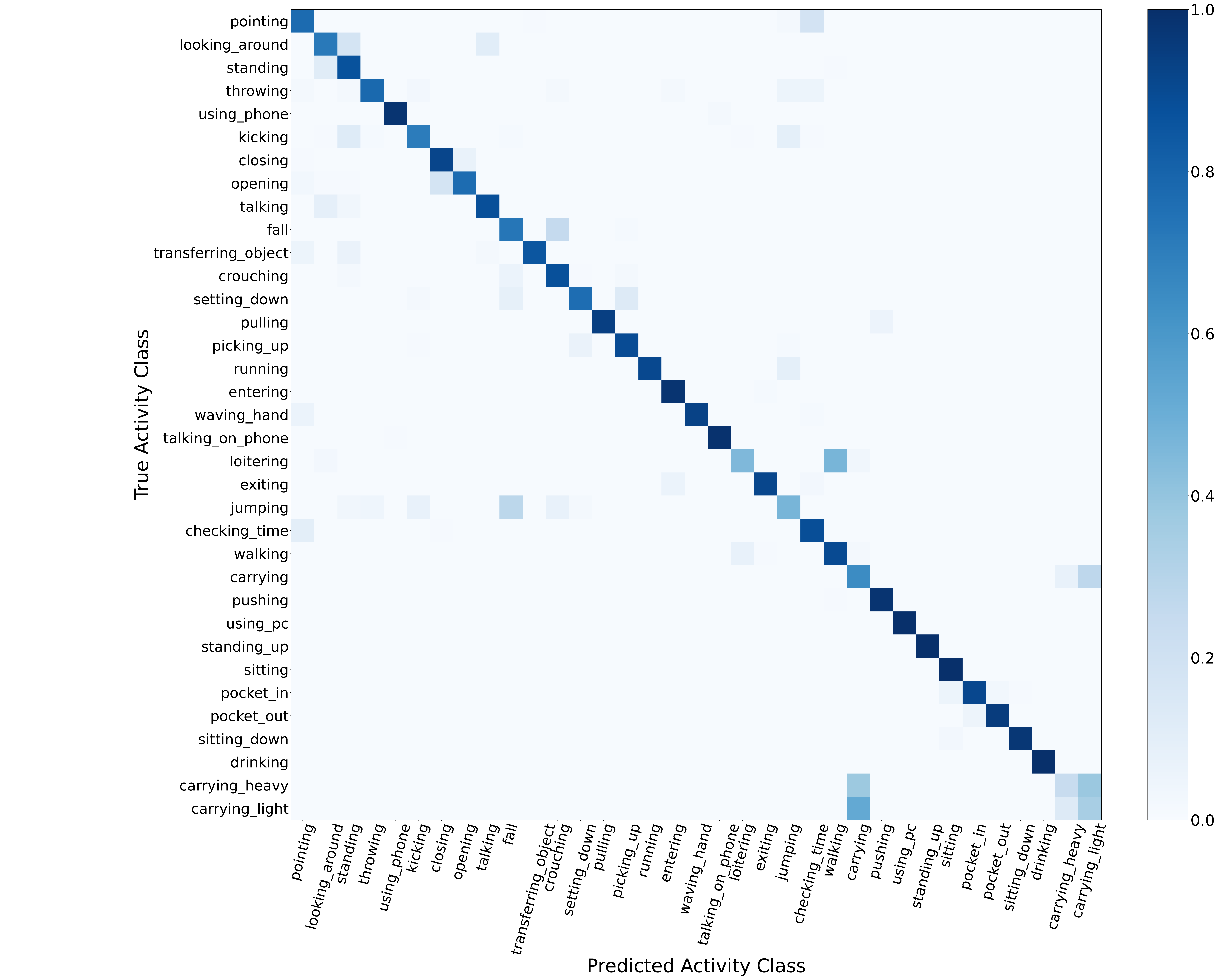}}
  \caption{Confusion matrices for the overall experiments on the UTD-MHAD dataset and the MMAct dataset. Sub-figures (a)-(d) are run on the UTD-MHAD dataset, where (a), (b) are under 50-50 subject setting using Inertial and Skeleton data and (c), (d) are under LOSO setting using RGB, Depth, Skeleton, and Inertial data. (e)-(h) are run on the MMAct dataset using Inertial and RGB data, where (e), (f) are under the cross-session setting and (g), (h) are under the cross-subject setting.}
  
\end{figure*}

For the MMAct dataset, we apply the cross-subject evaluation protocol and cross-session evaluation protocol and use the F1-Score as the evaluation metric. The experimental results are shown in Table \ref{tab:table_3} and Table \ref{tab:table_4}. The results indicate that the DMFT teacher model outperforms the other multi-modal approaches, except the predecessor MATN model by achieving 83.29\% under the cross-subject protocol and 91.62\% under the cross-session protocol. In the meanwhile, the DMFT student model achieves 82.54\% under the cross-subject setting and 91.09\% under the cross-session setting, which is slightly lower than the teacher model.

The overall results show that DMFT achieves competitive performance and outperforms several SOTA approaches. In general, the attentive models achieve better performance compared to the non-attentive models as the attention mechanism helps to extract the salient information. The results show that with the MSTT module, DMFT is able to extract the salient spatial and temporal features and achieves improved results compared to the non-attention approaches. While multi-modal approaches seem to improve the generalization ability, the area still lacks exploration. The other attention-based approaches apply a late-fusion method to fusee the multi-modal features, where the multi-modal features are concatenated after passing through the feature extraction module. However, DMFT applies the TMT module, which helps to conduct mid-fusion among the multi-modal features. Thus, the multi-modal streams share complementary information during the feature extraction which could improve the performance. The results show that the DMFT teacher model has shown good performance on the two datasets with different modality combinations. For example, the DMFT teacher model achieves 93.97\% accuracy which is 1.25\% higher than the SOTA MATN model under UTD-MHAD 50-50 subject setting. This is beneficial due to the privacy issue introduced when RGB features are used.

\begin{table*}[]
    \centering
        \caption{Performance comparison of the effeteness of the knowledge distillation method. R: RGB, S: Skeleton, D: Depth, I: Inertial. For results on the UTD-MHAD dataset, the top-1 accuracy is used, while for results on the MMAct dataset, the F1-score is used.}
    \label{tab:table_6}
    
    \begin{tabular}{cccccc}
        \hline
        { \bf Dataset } & { \bf Setting } & { \bf Modalities } & { \bf Teacher } & { \bf Student } & { \bf Student (KD) } \\
        \hline
        \multirow{4}{*}{UTD-MHAD} & 50-50 subject & S+I & 93.97 & 90.93 & 92.12 (1.19$\uparrow$) \\
        \cline{2-6}
         & \multirow{3}{*}{LOSO} & R+S & 93.06 & 89.96 & 90.26 (0.30$\uparrow$) \\
        \cline{3-6}
                              & & R+S+I & 98.20 & 96.41 & 96.52 (0.11$\uparrow$) \\
        \cline{3-6}
                              & & R+S+D+I & 97.52 & 96.27 & 96.53 (0.26$\uparrow$) \\
        \hline

        \multirow{2}{*}{MMAct} & cross-subject & I+R & 83.29 & 82.23 & 82.54 (0.31$\uparrow$) \\
        \cline{2-6}            
                               & cross-session & I+R & 91.62 & 90.90 & 91.08 (0.18$\uparrow$) \\
        \hline
    \end{tabular}
    
\end{table*}

\begin{table*}[]
    \centering
        \caption{Performance comparison of the efficiency of the knowledge distillation method. R: RGB, S: Skeleton, D: Depth, I: Inertial.}
    \label{tab:table_7}
    \begin{tabular}{cccccc}
        \hline
        { \bf Dataset } & { \bf Setting } & { \bf Modalities } & { \bf Model } & { \bf Training Time (s)} & { \bf Model Size (mb)} \\
        \hline
        \multirow{6}{*}{UTD-MHAD} & \multirow{3}{*}{LOSO} & \multirow{2}{*}{R+S} & Teacher & 6.44 & 1087 \\
        \cline{4-6}
                              & &  & Student (KD) & 2.32 & 364 \\
        \cline{3-6}
                              & & \multirow{2}{*}{R+S+I} & Teacher & 11.74 & 1188 \\
        \cline{4-6}
                              & &  & Student (KD) & 3.68 & 375 \\
        \cline{3-6}
                              & & \multirow{2}{*}{R+S+D+I} & Teacher & 16.08 & 1886 \\
        \cline{4-6}
                              & &  & Student (KD) & 5.88 & 577 \\
        \hline

        \multirow{4}{*}{MMAct} & \multirow{2}{*}{cross-subject} & \multirow{4}{*}{I+R} & Teacher & 84.74 & 1366 \\
        \cline{4-6}            
                               &  &  & Student (KD) & 44.39 & 321 \\
        \cline{2-2} \cline{4-6}            
                               & \multirow{2}{*}{cross-session} &  & Teacher & 81.05 & 2567 \\
        \cline{4-6}            
                               &  &  & Student (KD) & 43.20 & 588 \\
        \hline
    \end{tabular}
    
\end{table*}

\begin{table}[]
    \centering
        \caption{Performance comparison of the TFT tokens on the UTD-MHAD dataset.}
    \label{tab:table_5}
    \begin{tabular}{ccccc}
        \hline
        {\bf \# of TFT tokens}& 2 & 4 & 8 & 16\\ 
        \hline
        {\bf Accuracy (\%)} & 93.14& {\bf 93.72} & 92.65 & 92.93\\
        \hline
    \end{tabular}
    
\end{table}

From the results on the MMAct dataset we can see there is a gap in performance between the cross-subject setting and the cross-session setting, where for all the models, the performance under the cross-session setting is much higher. This is mainly because the cross-session setting is subject-dependent so that both the training set and the testing set share the characteristic of the same subject. When the model is deployed in a real-world situation, data of new subjects will be analyzed, rather than just the participants. In this case, developing models and evaluating their performance under a subject-dependent experimental protocol will lead to a completely different result, where the model’s performance will be overrated. This is in accordance with our motivation that we design subject-independent experimental protocols to evaluate our model’s performance and examine its generalization ability as this will be more accurate. More approaches should be explored to obtain better generalization ability. In the meantime, while there is still much space for improvement to conduct experiments on the MMAct dataset, the potential of the UTD-MHAD dataset seems to be well explored. This is because the UTD-MHAD only contains a small number of samples (861 clips). In the future, there is an urgent need of constructing comprehensive and large-scale multi-modal datasets.

\begin{table*}[]
    \centering
        \caption{Performance comparison of the efficiency of the knowledge distillation method. R: RGB, S: Skeleton, D: Depth, I: Inertial. For results on the UTD-MHAD dataset, the top-1 accuracy is used, while for results on the MMAct dataset, the F1-score is used.}
    \label{tab:table_8}
    \begin{tabular}{ccccccccc}
        \hline
        \multirow{2}{*}{ \bf Dataset } & \multirow{2}{*}{ \bf Setting } & \multirow{2}{*}{ \bf Modalities } & \multirow{2}{*}{ \bf Model } & \multicolumn{5}{c}{ \bf Result }  \\
        \cline{5-9}
                              & &  & &{ \bf R } & { \bf D }  & { \bf S } & { \bf I } & { \bf Overall }\\
        \hline
        \multirow{9}{*}{UTD-MHAD} & \multirow{9}{*}{LOSO} & \multirow{2}{*}{R+S} & Teacher & 61.02 & - & 91.77 & - & 93.06 \\
        \cline{4-9}
                              & &  & Student (Raw) & 56.46 & - & 89.57 & - & 89.96   \\
        \cline{4-9}
                              & &  & Student (KD) & 55.86 & - & 89.83 & - & 90.26   \\
        \cline{3-9}
                              & & \multirow{2}{*}{R+S+I} & Teacher & 59.05 & - & 91.16 & 79.19 & 98.20   \\
        \cline{4-9}
                              & &  & Student (Raw) & 53.80 & - & 88.59 & 78.07 & 96.41   \\
        \cline{4-9}                      
                              & &  & Student (KD) & 54.95 & - & 89.36 & 78.71 & 96.52   \\
        \cline{3-9}
                              & & \multirow{2}{*}{R+S+D+I} & Teacher & 57.03 & 39.75 & 90.22 & 78.27 & 97.52  \\
        \cline{4-9}
                              & &  & Student (Raw) & 53.82 & 31.04 & 88.04 & 77.74 & 96.27  \\
        \cline{4-9}                              
                              & &  & Student (KD) & 54.29 & 34.31 & 88.11 & 77.96 & 96.53  \\
        \hline

        \multirow{6}{*}{MMAct} & \multirow{3}{*}{cross-subject} & \multirow{6}{*}{R+I} & Teacher & 65.57 & - & - & 69.17 & 83.29  \\
        \cline{4-9}                
                               &  &  & Student (Raw) & 64.58 & - & - & 67.68 & 82.23  \\
        \cline{4-9} 
                               &  &  & Student (KD) & 65.59 & - & - & 67.82 & 82.54  \\
        \cline{2-2} \cline{4-9}            
                               & \multirow{3}{*}{cross-session} &  & Teacher & 74.28 & - & - & 82.75 & 91.62  \\
        \cline{4-9}            
                               &  &  & Student (Raw) & 74.27 & - & - & 80.25 & 90.90  \\
        \cline{4-9}            
                               &  &  & Student (KD) & 75.42 & - & - & 80.94 & 91.08  \\
        \hline
    \end{tabular}
    
\end{table*}

\subsection{Impact of the Temporal Mid-fusion Tokens}

In the DMFT teacher model, we use the TMT tokens to conduct mid-fusion among the multi-modal temporal features. In this section, we conduct an ablation study to evaluate if the number of TMT tokens would have much influence on the mid-fusion process. We conduct experiments on the UTD-MHAD dataset using the 50-50 subject setting. The only difference when constructing the models is using different numbers of TMT tokens. The results are shown in table \ref{tab:table_5}.

The results show that using more TMT tokens would not have a significant positive influence on the model’s performance. This aligns with the work BMT’s conclusion \cite{nagrani2021attention} that using a small number of fusion tokens is enough to share the common information among the multi-modal features. Thus we use 4 TMT tokens as this would reduce the computation cost whilst achieving better performance.

\subsection{Effeteness of Knowledge Distillation}

In this section, we conduct an ablation study to evaluate the knowledge distillation method’s influence to improve the student network’s performance. For each experimental setting, we train a raw student network without applying the knowledge distillation step. The results are shown in table \ref{tab:table_6}. For the experiments on the UTD-MHAD dataset, we use Top-1 accuracy, while for the experiments on the MMAct dataset, we use the F1-score.

The results show that there is an improvement in terms of performance when a teacher network is used to train the student network. The maximum improvement is 1.19\% when the 50-50 setting is used on the UTD-MHAD dataset. For the MMAct dataset, there is an improvement of 0.31\% when the cross-subject setting is applied. While there is a minor improvement (0.18\%) under the cross-session setting, the evaluation protocol is subject-dependent so it cannot reflect the situation in the real-world condition. The results are in accordance with our motivation that by applying the Knowledge Distillation approach, we can transfer the knowledge from a complex teacher model to a smaller student network to improve its performance.

\subsection{Efficiency of Knowledge Distillation}

In this section, we present a comparative evaluation to demonstrate the efficiency of utilizing the Knowledge Distillation method to train the student models. The results are shown in table \ref{tab:table_7}, which includes the required training time per epoch and the saved model size for the teacher model, the student model, and the KD student model. Experiments on the UTD-MHAD dataset are run on an NVIDIA RTX 3090 GPU, and experiments on the MMAct dataset are run on an NVIDIA A40 GPU.

The results show that by applying the KD method, both the training time and the model size are significantly reduced, which is presented across different settings. This supports our motivation that applying the KD method to train the student model could reduce the time and space cost of the model. As the hardware devices may be limited in real-world scenarios, our approach would be beneficial for real-life deployment.

\subsection{Impact of multi-modal Learning}

In this section, we conduct a further study to evaluate DMFT’s performance in multi-modal learning. We conduct the study on both the UTD-MHAD dataset and the MMAct dataset under different experimental protocols and modality combinations. For each experimental setting, we train 3 models, the teacher network, the raw student network, and the student network with knowledge distillation. For each modality combination, we present the performance comparison among each modality stream’s output and the overall output. The results are presented in table \ref{tab:table_8}. Also, we present Figure \ref{Fig.5} and Figure \ref{Fig.6}, which show a more detailed performance evaluation across each activity. For the experiments on the UTD-MHAD dataset, we use Top-1 accuracy, while for the experiments on the MMAct dataset, we use the F-1 score. The results show that DMFT can capture the complementary information from each modality and make well use of the salient features, thus producing enhanced results.

\begin{figure*}[]
\centering 
\includegraphics[width=0.8\textwidth]{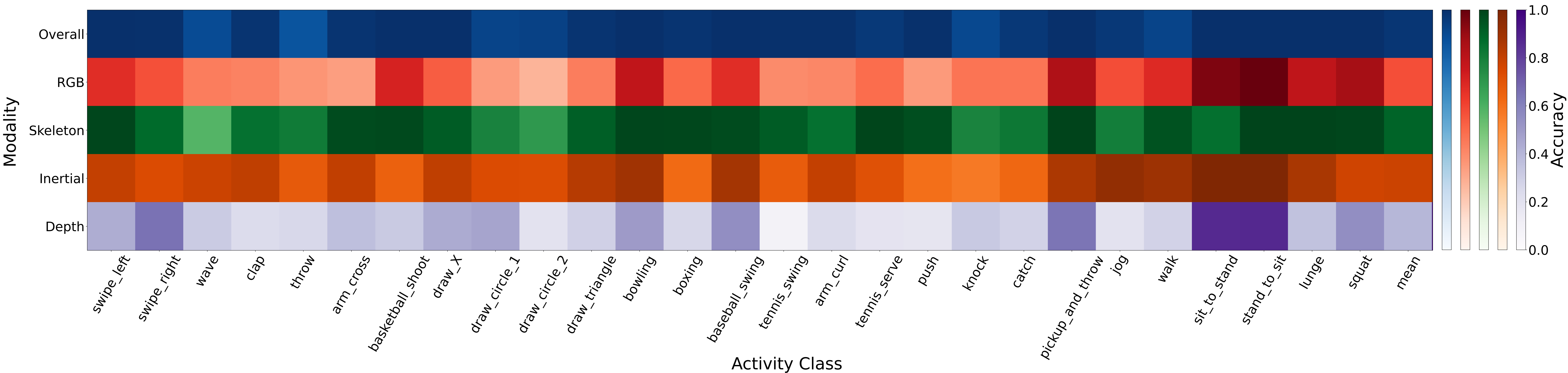} 
\caption{Performance comparison on the contribution of modalities on the UTD-MHAD dataset (Top-1 Accuracy)} 
\label{Fig.5} 
\end{figure*}

\begin{figure*}[]
\centering 
\includegraphics[width=0.8\textwidth]{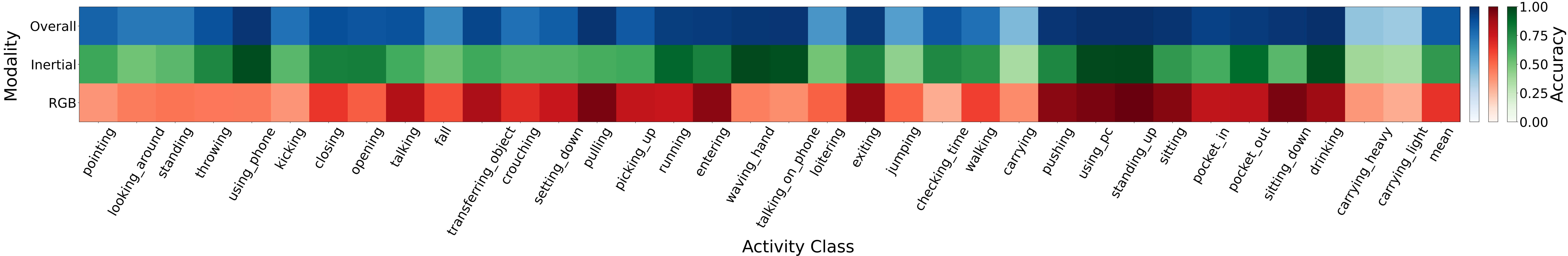} 
\caption{Performance comparison on the contribution of modalities on the MMAct dataset (F1-Score)} 
\label{Fig.6} 
\end{figure*}

One of the advantages of multi-modal learning is to make use of complementary information to produce more accurate and robust results. The results in table \ref{tab:table_8} show that for all the experimental settings, the overall result achieves a better performance. While the performance of each modality stream may vary, by aggregating the output of each modality, the model is able to capture the salient modality-specific features. In this case, even if one modality input failed, the model would still be able to conduct feature extraction using the other modalities and capture the complementary information, to produce robust predictions. For example, in Figure \ref{Fig.5}, while the skeleton stream performs better than the other three modalities for class 1 (swipe left), the inertial stream outperforms the other modalities for class 3 (wave). However, after aggregating the information of all the modality streams, the overall prediction outperforms each modality stream. Also, Figure \ref{Fig.6}, while both the RGB and the Inertial streams have a performance lower than 80\% for class 30 (pocket in), the overall prediction achieves an F1-score over 90\%. As a result, applying multi-modal learning could produce more robust results which is beneficial in the real-world deployment as the signal transmission may be affected. 

\section{Conclusion}
The main objective of our work is to develop an effective and efficient multi-modal human activity recognition approach, which can be deployed in resource-limited environments and generalized in subject-independent settings. We present DMFT, a knowledge distillation based attentive approach that conducts mid-fusion among the multi-modal features to resolve the multi-modal human activity recognition task. We first encode the multi-modal data through the unified representation learning layer. Then we apply the Multi-modal Temporal Mid-Fusion Transformer Network to extract the salient spatial-temporal features of each modality and conduct temporal mid-fusion to further extract and fuse the multi-modal features.
We also apply a knowledge distillation method and use the teacher network to train a simpler student network, which improves the performance whilst reducing the computation and space cost. We conduct comprehensive experiments on two public multi-modal datasets, UTD-MHAD and MMAct under different experimental settings to evaluate DMFT’s performance. The experimental results show that our model can make use of the salient multi-modal features and produce competitive results while being able to achieve improved and robust performance in a limited environment. In the future, we plan to develop effective, efficient, and robust human activity recognition models that can better resolve the inter-subject variation challenge in a multi-modal human activity recognition scenario.

\section*{Declarations}
This work was supported by the Cooperative Research Centres Projects (CRCP) Grants "DeepIoT" project. 




\bibliographystyle{elsarticle-num-names} 
\bibliography{ref}

\begin{thebibliography}{55}
\expandafter\ifx\csname natexlab\endcsname\relax\def\natexlab#1{#1}\fi
\providecommand{\url}[1]{\texttt{#1}}
\providecommand{\href}[2]{#2}
\providecommand{\path}[1]{#1}
\providecommand{\DOIprefix}{doi:}
\providecommand{\ArXivprefix}{arXiv:}
\providecommand{\URLprefix}{URL: }
\providecommand{\Pubmedprefix}{pmid:}
\providecommand{\doi}[1]{\href{http://dx.doi.org/#1}{\path{#1}}}
\providecommand{\Pubmed}[1]{\href{pmid:#1}{\path{#1}}}
\providecommand{\bibinfo}[2]{#2}
\ifx\xfnm\relax \def\xfnm[#1]{\unskip,\space#1}\fi
\bibitem[{Chen et~al.(2015)Chen, Jafari, and Kehtarnavaz}]{chen2015utd}
\bibinfo{author}{C.~Chen}, \bibinfo{author}{R.~Jafari},
  \bibinfo{author}{N.~Kehtarnavaz},
\newblock \bibinfo{title}{Utd-mhad: A multimodal dataset for human action
  recognition utilizing a depth camera and a wearable inertial sensor},
\newblock in: \bibinfo{booktitle}{2015 IEEE International Conference on Image
  Processing (ICIP)}, \bibinfo{organization}{IEEE}, \bibinfo{year}{2015}, pp.
  \bibinfo{pages}{168--172}.
\bibitem[{Kong et~al.(2019)Kong, Wu, Deng, Klinkigt, Tong, and
  Murakami}]{kong2019mmact}
\bibinfo{author}{Q.~Kong}, \bibinfo{author}{Z.~Wu}, \bibinfo{author}{Z.~Deng},
  \bibinfo{author}{M.~Klinkigt}, \bibinfo{author}{B.~Tong},
  \bibinfo{author}{T.~Murakami},
\newblock \bibinfo{title}{Mmact: A large-scale dataset for cross modal human
  action understanding},
\newblock in: \bibinfo{booktitle}{Proceedings of the IEEE/CVF International
  Conference on Computer Vision}, \bibinfo{year}{2019}, pp.
  \bibinfo{pages}{8658--8667}.
\bibitem[{Brand and Hertzmann(2000)}]{brand2000style}
\bibinfo{author}{M.~Brand}, \bibinfo{author}{A.~Hertzmann},
\newblock \bibinfo{title}{Style machines},
\newblock in: \bibinfo{booktitle}{Proceedings of the 27th Annual Conference on
  Computer Graphics and Interactive Techniques}, SIGGRAPH '00,
  \bibinfo{publisher}{ACM Press/Addison-Wesley Publishing Co.},
  \bibinfo{address}{USA}, \bibinfo{year}{2000}, p. \bibinfo{pages}{183–192}.
  \URLprefix \url{https://doi.org/10.1145/344779.344865}.
  \DOIprefix\doi{10.1145/344779.344865}.
\bibitem[{Pavlovic et~al.(2000)Pavlovic, Rehg, and
  MacCormick}]{pavlovic2000learning}
\bibinfo{author}{V.~Pavlovic}, \bibinfo{author}{J.~M. Rehg},
  \bibinfo{author}{J.~MacCormick},
\newblock \bibinfo{title}{Learning switching linear models of human motion},
\newblock in: \bibinfo{booktitle}{Proceedings of the 13th International
  Conference on Neural Information Processing Systems}, NIPS'00,
  \bibinfo{publisher}{MIT Press}, \bibinfo{address}{Cambridge, MA, USA},
  \bibinfo{year}{2000}, p. \bibinfo{pages}{942–948}.
\bibitem[{Urtasun et~al.(2008)Urtasun, Fleet, Geiger, Popovi\'{c}, Darrell, and
  Lawrence}]{urtasun2008topo}
\bibinfo{author}{R.~Urtasun}, \bibinfo{author}{D.~J. Fleet},
  \bibinfo{author}{A.~Geiger}, \bibinfo{author}{J.~Popovi\'{c}},
  \bibinfo{author}{T.~J. Darrell}, \bibinfo{author}{N.~D. Lawrence},
\newblock \bibinfo{title}{Topologically-constrained latent variable models},
\newblock in: \bibinfo{booktitle}{Proceedings of the 25th International
  Conference on Machine Learning}, ICML '08, \bibinfo{publisher}{Association
  for Computing Machinery}, \bibinfo{address}{New York, NY, USA},
  \bibinfo{year}{2008}, p. \bibinfo{pages}{1080–1087}. \URLprefix
  \url{https://doi.org/10.1145/1390156.1390292}.
  \DOIprefix\doi{10.1145/1390156.1390292}.
\bibitem[{Wang et~al.(2008)Wang, Fleet, and Hertzmann}]{wang2008gaussian}
\bibinfo{author}{J.~M. Wang}, \bibinfo{author}{D.~J. Fleet},
  \bibinfo{author}{A.~Hertzmann},
\newblock \bibinfo{title}{Gaussian process dynamical models for human motion},
\newblock \bibinfo{journal}{IEEE Transactions on Pattern Analysis and Machine
  Intelligence} \bibinfo{volume}{30} (\bibinfo{year}{2008})
  \bibinfo{pages}{283--298}. \DOIprefix\doi{10.1109/TPAMI.2007.1167}.
\bibitem[{Akhter et~al.(2012)Akhter, Simon, Khan, Matthews, and
  Sheikh}]{akhter2012bilinear}
\bibinfo{author}{I.~Akhter}, \bibinfo{author}{T.~Simon},
  \bibinfo{author}{S.~Khan}, \bibinfo{author}{I.~Matthews},
  \bibinfo{author}{Y.~Sheikh},
\newblock \bibinfo{title}{Bilinear spatiotemporal basis models},
\newblock \bibinfo{journal}{ACM Trans. Graph.} \bibinfo{volume}{31}
  (\bibinfo{year}{2012}). \URLprefix
  \url{https://doi.org/10.1145/2159516.2159523}.
  \DOIprefix\doi{10.1145/2159516.2159523}.
\bibitem[{Sutskever et~al.(2008)Sutskever, Hinton, and
  Taylor}]{sutskever2008the}
\bibinfo{author}{I.~Sutskever}, \bibinfo{author}{G.~Hinton},
  \bibinfo{author}{G.~Taylor},
\newblock \bibinfo{title}{The recurrent temporal restricted boltzmann machine},
\newblock in: \bibinfo{booktitle}{Proceedings of the 21st International
  Conference on Neural Information Processing Systems}, NIPS'08,
  \bibinfo{publisher}{Curran Associates Inc.}, \bibinfo{address}{Red Hook, NY,
  USA}, \bibinfo{year}{2008}, p. \bibinfo{pages}{1601–1608}.
\bibitem[{Taylor et~al.(2010)Taylor, Sigal, Fleet, and
  Hinton}]{taylor2010dynamical}
\bibinfo{author}{G.~W. Taylor}, \bibinfo{author}{L.~Sigal},
  \bibinfo{author}{D.~J. Fleet}, \bibinfo{author}{G.~E. Hinton},
\newblock \bibinfo{title}{Dynamical binary latent variable models for 3d human
  pose tracking},
\newblock in: \bibinfo{booktitle}{2010 IEEE Computer Society Conference on
  Computer Vision and Pattern Recognition}, \bibinfo{year}{2010}, pp.
  \bibinfo{pages}{631--638}. \DOIprefix\doi{10.1109/CVPR.2010.5540157}.
\bibitem[{Fragkiadaki et~al.(2015)Fragkiadaki, Levine, Felsen, and
  Malik}]{fragkiadaki2015recurrent}
\bibinfo{author}{K.~Fragkiadaki}, \bibinfo{author}{S.~Levine},
  \bibinfo{author}{P.~Felsen}, \bibinfo{author}{J.~Malik},
\newblock \bibinfo{title}{Recurrent network models for human dynamics},
\newblock in: \bibinfo{booktitle}{2015 IEEE International Conference on
  Computer Vision (ICCV)}, \bibinfo{year}{2015}, pp.
  \bibinfo{pages}{4346--4354}. \DOIprefix\doi{10.1109/ICCV.2015.494}.
\bibitem[{Wang et~al.(2016)Wang, Gao, Song, and Shen}]{wang2016beyond}
\bibinfo{author}{X.~Wang}, \bibinfo{author}{L.~Gao}, \bibinfo{author}{J.~Song},
  \bibinfo{author}{H.~Shen},
\newblock \bibinfo{title}{Beyond frame-level cnn: saliency-aware 3-d cnn with
  lstm for video action recognition},
\newblock \bibinfo{journal}{IEEE Signal Processing Letters}
  \bibinfo{volume}{24} (\bibinfo{year}{2016}) \bibinfo{pages}{510--514}.
\bibitem[{Martinez et~al.(2017)Martinez, Black, and Romero}]{martinez2017human}
\bibinfo{author}{J.~Martinez}, \bibinfo{author}{M.~J. Black},
  \bibinfo{author}{J.~Romero},
\newblock \bibinfo{title}{On human motion prediction using recurrent neural
  networks},
\newblock in: \bibinfo{booktitle}{Proceedings of the IEEE conference on
  Computer Vision and Pattern Recognition}, \bibinfo{year}{2017}, pp.
  \bibinfo{pages}{2891--2900}.
\bibitem[{Lee et~al.(2017)Lee, Jung, and Tani}]{lee2017recognition}
\bibinfo{author}{H.~Lee}, \bibinfo{author}{M.~Jung}, \bibinfo{author}{J.~Tani},
\newblock \bibinfo{title}{Recognition of visually perceived compositional human
  actions by multiple spatio-temporal scales recurrent neural networks},
\newblock \bibinfo{journal}{IEEE Transactions on Cognitive and Developmental
  Systems} \bibinfo{volume}{10} (\bibinfo{year}{2017})
  \bibinfo{pages}{1058--1069}.
\bibitem[{Gui et~al.(2018)Gui, Wang, Liang, and Moura}]{gui2018adversarial}
\bibinfo{author}{L.-Y. Gui}, \bibinfo{author}{Y.-X. Wang},
  \bibinfo{author}{X.~Liang}, \bibinfo{author}{J.~M. Moura},
\newblock \bibinfo{title}{Adversarial geometry-aware human motion prediction},
\newblock in: \bibinfo{booktitle}{Proceedings of the European Conference on
  Computer Vision (ECCV)}, \bibinfo{year}{2018}, pp. \bibinfo{pages}{786--803}.
\bibitem[{Guo et~al.(2018)Guo, Chou, Huang, Song, Yeung, and
  Fei-Fei}]{guo2018neural}
\bibinfo{author}{M.~Guo}, \bibinfo{author}{E.~Chou}, \bibinfo{author}{D.-A.
  Huang}, \bibinfo{author}{S.~Song}, \bibinfo{author}{S.~Yeung},
  \bibinfo{author}{L.~Fei-Fei},
\newblock \bibinfo{title}{Neural graph matching networks for fewshot 3d action
  recognition},
\newblock in: \bibinfo{booktitle}{Proceedings of the European Conference on
  Computer Vision (ECCV)}, \bibinfo{year}{2018}, pp. \bibinfo{pages}{653--669}.
\bibitem[{Fan et~al.(2013)Fan, Wang, and Wang}]{fan2013human}
\bibinfo{author}{L.~Fan}, \bibinfo{author}{Z.~Wang}, \bibinfo{author}{H.~Wang},
\newblock \bibinfo{title}{Human activity recognition model based on decision
  tree},
\newblock in: \bibinfo{booktitle}{2013 International Conference on Advanced
  Cloud and Big Data}, \bibinfo{organization}{IEEE}, \bibinfo{year}{2013}, pp.
  \bibinfo{pages}{64--68}.
\bibitem[{Paul and George(2015)}]{paul2015effective}
\bibinfo{author}{P.~Paul}, \bibinfo{author}{T.~George},
\newblock \bibinfo{title}{An effective approach for human activity recognition
  on smartphone},
\newblock in: \bibinfo{booktitle}{2015 IEEE International Conference on
  Engineering and Technology (ICETECH)}, \bibinfo{organization}{IEEE},
  \bibinfo{year}{2015}, pp. \bibinfo{pages}{1--3}.
\bibitem[{Chathuramali and Rodrigo(2012)}]{chathuramali2012faster}
\bibinfo{author}{K.~M. Chathuramali}, \bibinfo{author}{R.~Rodrigo},
\newblock \bibinfo{title}{Faster human activity recognition with svm},
\newblock in: \bibinfo{booktitle}{International Conference on Advances in ICT
  for Emerging Regions (ICTer2012)}, \bibinfo{organization}{IEEE},
  \bibinfo{year}{2012}, pp. \bibinfo{pages}{197--203}.
\bibitem[{Liu et~al.(2015)Liu, Peng, Liu, and Huang}]{liu2015sensor}
\bibinfo{author}{L.~Liu}, \bibinfo{author}{Y.~Peng}, \bibinfo{author}{M.~Liu},
  \bibinfo{author}{Z.~Huang},
\newblock \bibinfo{title}{Sensor-based human activity recognition system with a
  multilayered model using time series shapelets},
\newblock \bibinfo{journal}{Knowledge-Based Systems} \bibinfo{volume}{90}
  (\bibinfo{year}{2015}) \bibinfo{pages}{138--152}.
\bibitem[{Fallmann and Kropf(2016)}]{fallmann2016human}
\bibinfo{author}{S.~Fallmann}, \bibinfo{author}{J.~Kropf},
\newblock \bibinfo{title}{Human activity recognition of continuous data using
  hidden markov models and the aspect of including discrete data},
\newblock in: \bibinfo{booktitle}{2016 Intl IEEE Conferences on Ubiquitous
  Intelligence \& Computing, Advanced and Trusted Computing, Scalable Computing
  and Communications, Cloud and Big Data Computing, Internet of People, and
  Smart World Congress (UIC/ATC/ScalCom/CBDCom/IoP/SmartWorld)},
  \bibinfo{organization}{IEEE}, \bibinfo{year}{2016}, pp.
  \bibinfo{pages}{121--126}.
\bibitem[{Kabir et~al.(2016)Kabir, Hoque, Thapa, and Yang}]{kabir2016two}
\bibinfo{author}{M.~H. Kabir}, \bibinfo{author}{M.~R. Hoque},
  \bibinfo{author}{K.~Thapa}, \bibinfo{author}{S.-H. Yang},
\newblock \bibinfo{title}{Two-layer hidden markov model for human activity
  recognition in home environments},
\newblock \bibinfo{journal}{International Journal of Distributed Sensor
  Networks} \bibinfo{volume}{12} (\bibinfo{year}{2016})
  \bibinfo{pages}{4560365}.
\bibitem[{Jiang and Yin(2015)}]{jiang2015human}
\bibinfo{author}{W.~Jiang}, \bibinfo{author}{Z.~Yin},
\newblock \bibinfo{title}{Human activity recognition using wearable sensors by
  deep convolutional neural networks},
\newblock in: \bibinfo{booktitle}{Proceedings of the 23rd ACM international
  conference on Multimedia}, \bibinfo{year}{2015}, pp.
  \bibinfo{pages}{1307--1310}.
\bibitem[{Yang et~al.(2015)Yang, Nguyen, San, Li, and
  Krishnaswamy}]{yang2015deep}
\bibinfo{author}{J.~B. Yang}, \bibinfo{author}{M.~N. Nguyen},
  \bibinfo{author}{P.~P. San}, \bibinfo{author}{X.~L. Li},
  \bibinfo{author}{S.~Krishnaswamy},
\newblock \bibinfo{title}{Deep convolutional neural networks on multichannel
  time series for human activity recognition},
\newblock in: \bibinfo{booktitle}{Proceedings of the Twenty-Fourth
  International Conference on Artificial Intelligence}, IJCAI'15,
  \bibinfo{publisher}{AAAI Press}, \bibinfo{year}{2015}, p.
  \bibinfo{pages}{3995–4001}.
\bibitem[{Peng et~al.(2018)Peng, Chen, Ye, and Zhang}]{peng2018aroma}
\bibinfo{author}{L.~Peng}, \bibinfo{author}{L.~Chen}, \bibinfo{author}{Z.~Ye},
  \bibinfo{author}{Y.~Zhang},
\newblock \bibinfo{title}{Aroma: A deep multi-task learning based simple and
  complex human activity recognition method using wearable sensors},
\newblock \bibinfo{journal}{Proceedings of the ACM on Interactive, Mobile,
  Wearable and Ubiquitous Technologies} \bibinfo{volume}{2}
  (\bibinfo{year}{2018}) \bibinfo{pages}{1--16}.
\bibitem[{Chen et~al.(2019)Chen, Yao, Zhang, Guo, and Yu}]{chen2019multi}
\bibinfo{author}{K.~Chen}, \bibinfo{author}{L.~Yao},
  \bibinfo{author}{D.~Zhang}, \bibinfo{author}{B.~Guo},
  \bibinfo{author}{Z.~Yu},
\newblock \bibinfo{title}{Multi-agent attentional activity recognition},
\newblock in: \bibinfo{booktitle}{Proceedings of the Twenty-Eighth
  International Joint Conference on Artificial Intelligence, {IJCAI-19}},
  \bibinfo{publisher}{International Joint Conferences on Artificial
  Intelligence Organization}, \bibinfo{year}{2019}, pp.
  \bibinfo{pages}{1344--1350}. \URLprefix
  \url{https://doi.org/10.24963/ijcai.2019/186}.
  \DOIprefix\doi{10.24963/ijcai.2019/186}.
\bibitem[{Xue et~al.(2020)Xue, Jiang, Miao, Ma, Wang, Yuan, Yao, Zhang, and
  Su}]{xue2020deepmv}
\bibinfo{author}{H.~Xue}, \bibinfo{author}{W.~Jiang},
  \bibinfo{author}{C.~Miao}, \bibinfo{author}{F.~Ma},
  \bibinfo{author}{S.~Wang}, \bibinfo{author}{Y.~Yuan},
  \bibinfo{author}{S.~Yao}, \bibinfo{author}{A.~Zhang},
  \bibinfo{author}{L.~Su},
\newblock \bibinfo{title}{Deepmv: Multi-view deep learning for device-free
  human activity recognition},
\newblock \bibinfo{journal}{Proceedings of the ACM on Interactive, Mobile,
  Wearable and Ubiquitous Technologies} \bibinfo{volume}{4}
  (\bibinfo{year}{2020}) \bibinfo{pages}{1--26}.
\bibitem[{Bai et~al.(2020)Bai, Yao, Wang, Kanhere, Guo, and
  Yu}]{bai2020adversarial}
\bibinfo{author}{L.~Bai}, \bibinfo{author}{L.~Yao}, \bibinfo{author}{X.~Wang},
  \bibinfo{author}{S.~S. Kanhere}, \bibinfo{author}{B.~Guo},
  \bibinfo{author}{Z.~Yu},
\newblock \bibinfo{title}{Adversarial multi-view networks for activity
  recognition},
\newblock \bibinfo{journal}{Proceedings of the ACM on Interactive, Mobile,
  Wearable and Ubiquitous Technologies} \bibinfo{volume}{4}
  (\bibinfo{year}{2020}) \bibinfo{pages}{1--22}.
\bibitem[{Guan and Pl{\"o}tz(2017)}]{guan2017ensembles}
\bibinfo{author}{Y.~Guan}, \bibinfo{author}{T.~Pl{\"o}tz},
\newblock \bibinfo{title}{Ensembles of deep lstm learners for activity
  recognition using wearables},
\newblock \bibinfo{journal}{Proceedings of the ACM on Interactive, Mobile,
  Wearable and Ubiquitous Technologies} \bibinfo{volume}{1}
  (\bibinfo{year}{2017}) \bibinfo{pages}{1--28}.
\bibitem[{Murahari and Pl{\"o}tz(2018)}]{murahari2018attention}
\bibinfo{author}{V.~S. Murahari}, \bibinfo{author}{T.~Pl{\"o}tz},
\newblock \bibinfo{title}{On attention models for human activity recognition},
\newblock in: \bibinfo{booktitle}{Proceedings of the 2018 ACM international
  symposium on wearable computers}, \bibinfo{year}{2018}, pp.
  \bibinfo{pages}{100--103}.
\bibitem[{Zeng et~al.(2018)Zeng, Gao, Yu, Mengshoel, Langseth, Lane, and
  Liu}]{zeng2018understanding}
\bibinfo{author}{M.~Zeng}, \bibinfo{author}{H.~Gao}, \bibinfo{author}{T.~Yu},
  \bibinfo{author}{O.~J. Mengshoel}, \bibinfo{author}{H.~Langseth},
  \bibinfo{author}{I.~Lane}, \bibinfo{author}{X.~Liu},
\newblock \bibinfo{title}{Understanding and improving recurrent networks for
  human activity recognition by continuous attention},
\newblock in: \bibinfo{booktitle}{Proceedings of the 2018 ACM International
  Symposium on Wearable Computers}, \bibinfo{year}{2018}, pp.
  \bibinfo{pages}{56--63}.
\bibitem[{Bock et~al.(2021)Bock, H{\"o}lzemann, Moeller, and
  Van~Laerhoven}]{bock2021improving}
\bibinfo{author}{M.~Bock}, \bibinfo{author}{A.~H{\"o}lzemann},
  \bibinfo{author}{M.~Moeller}, \bibinfo{author}{K.~Van~Laerhoven},
\newblock \bibinfo{title}{Improving deep learning for har with shallow lstms},
\newblock in: \bibinfo{booktitle}{2021 International Symposium on Wearable
  Computers}, \bibinfo{year}{2021}, pp. \bibinfo{pages}{7--12}.
\bibitem[{Wu et~al.(2023)Wu, Zhang, Li, Shang, Han, Geng, and
  Pan}]{wu2023pedal}
\bibinfo{author}{H.~Wu}, \bibinfo{author}{Z.~Zhang}, \bibinfo{author}{X.~Li},
  \bibinfo{author}{K.~Shang}, \bibinfo{author}{Y.~Han},
  \bibinfo{author}{Z.~Geng}, \bibinfo{author}{T.~Pan},
\newblock \bibinfo{title}{A novel pedal musculoskeletal response based on
  differential spatio-temporal lstm for human activity recognition},
\newblock \bibinfo{journal}{Knowledge-Based Systems} \bibinfo{volume}{261}
  (\bibinfo{year}{2023}) \bibinfo{pages}{110187}. \URLprefix
  \url{https://www.sciencedirect.com/science/article/pii/S0950705122012837}.
  \DOIprefix\doi{https://doi.org/10.1016/j.knosys.2022.110187}.
\bibitem[{Guo et~al.(2016)Guo, Chen, Peng, and Chen}]{guo2016wearable}
\bibinfo{author}{H.~Guo}, \bibinfo{author}{L.~Chen}, \bibinfo{author}{L.~Peng},
  \bibinfo{author}{G.~Chen},
\newblock \bibinfo{title}{Wearable sensor based multimodal human activity
  recognition exploiting the diversity of classifier ensemble},
\newblock in: \bibinfo{booktitle}{Proceedings of the 2016 ACM International
  Joint Conference on Pervasive and Ubiquitous Computing},
  \bibinfo{year}{2016}, pp. \bibinfo{pages}{1112--1123}.
\bibitem[{Yao et~al.(2018)Yao, Sheng, Benatallah, Dustdar, Wang, Shemshadi, and
  Kanhere}]{yao2018wits}
\bibinfo{author}{L.~Yao}, \bibinfo{author}{Q.~Z. Sheng},
  \bibinfo{author}{B.~Benatallah}, \bibinfo{author}{S.~Dustdar},
  \bibinfo{author}{X.~Wang}, \bibinfo{author}{A.~Shemshadi},
  \bibinfo{author}{S.~S. Kanhere},
\newblock \bibinfo{title}{Wits: an iot-endowed computational framework for
  activity recognition in personalized smart homes},
\newblock \bibinfo{journal}{Computing} \bibinfo{volume}{100}
  (\bibinfo{year}{2018}) \bibinfo{pages}{369--385}.
\bibitem[{Memmesheimer et~al.(2020)Memmesheimer, Theisen, and
  Paulus}]{memmesheimer2020gimme}
\bibinfo{author}{R.~Memmesheimer}, \bibinfo{author}{N.~Theisen},
  \bibinfo{author}{D.~Paulus},
\newblock \bibinfo{title}{Gimme signals: Discriminative signal encoding for
  multimodal activity recognition},
\newblock in: \bibinfo{booktitle}{2020 IEEE/RSJ International Conference on
  Intelligent Robots and Systems (IROS)}, \bibinfo{organization}{IEEE},
  \bibinfo{year}{2020}, pp. \bibinfo{pages}{10394--10401}.
\bibitem[{Islam and Iqbal(2020)}]{islam2020hamlet}
\bibinfo{author}{M.~M. Islam}, \bibinfo{author}{T.~Iqbal},
\newblock \bibinfo{title}{Hamlet: A hierarchical multimodal attention-based
  human activity recognition algorithm},
\newblock in: \bibinfo{booktitle}{2020 IEEE/RSJ International Conference on
  Intelligent Robots and Systems (IROS)}, \bibinfo{organization}{IEEE},
  \bibinfo{year}{2020}, pp. \bibinfo{pages}{10285--10292}.
\bibitem[{Islam and Iqbal(2021)}]{islam2021multi}
\bibinfo{author}{M.~M. Islam}, \bibinfo{author}{T.~Iqbal},
\newblock \bibinfo{title}{Multi-gat: A graphical attention-based hierarchical
  multimodal representation learning approach for human activity recognition},
\newblock \bibinfo{journal}{IEEE Robotics and Automation Letters}
  \bibinfo{volume}{6} (\bibinfo{year}{2021}) \bibinfo{pages}{1729--1736}.
\bibitem[{Li et~al.(2022)Li, Yao, Li, Wang, and Sammut}]{li2022multi}
\bibinfo{author}{J.~Li}, \bibinfo{author}{L.~Yao}, \bibinfo{author}{B.~Li},
  \bibinfo{author}{X.~Wang}, \bibinfo{author}{C.~Sammut},
\newblock \bibinfo{title}{Multi-agent transformer networks for multimodal human
  activity recognition},
\newblock in: \bibinfo{booktitle}{Proceedings of the 31st ACM International
  Conference on Information \& Knowledge Management}, \bibinfo{year}{2022}, pp.
  \bibinfo{pages}{1135--1145}.
\bibitem[{Chen et~al.(2019)Chen, Yao, Zhang, Wang, Chang, and
  Nie}]{chen2019semisupervised}
\bibinfo{author}{K.~Chen}, \bibinfo{author}{L.~Yao},
  \bibinfo{author}{D.~Zhang}, \bibinfo{author}{X.~Wang},
  \bibinfo{author}{X.~Chang}, \bibinfo{author}{F.~Nie},
\newblock \bibinfo{title}{A semisupervised recurrent convolutional attention
  model for human activity recognition},
\newblock \bibinfo{journal}{IEEE transactions on neural networks and learning
  systems} \bibinfo{volume}{31} (\bibinfo{year}{2019})
  \bibinfo{pages}{1747--1756}.
\bibitem[{Mnih et~al.(2014)Mnih, Heess, Graves et~al.}]{mnih2014recurrent}
\bibinfo{author}{V.~Mnih}, \bibinfo{author}{N.~Heess},
  \bibinfo{author}{A.~Graves}, et~al.,
\newblock \bibinfo{title}{Recurrent models of visual attention},
\newblock \bibinfo{journal}{Advances in neural information processing systems}
  \bibinfo{volume}{27} (\bibinfo{year}{2014}).
\bibitem[{Long et~al.(2018)Long, Gan, De~Melo, Liu, Li, Li, and
  Wen}]{long2018multimodal}
\bibinfo{author}{X.~Long}, \bibinfo{author}{C.~Gan},
  \bibinfo{author}{G.~De~Melo}, \bibinfo{author}{X.~Liu},
  \bibinfo{author}{Y.~Li}, \bibinfo{author}{F.~Li}, \bibinfo{author}{S.~Wen},
\newblock \bibinfo{title}{Multimodal keyless attention fusion for video
  classification},
\newblock in: \bibinfo{booktitle}{Thirty-Second AAAI Conference on Artificial
  Intelligence}, \bibinfo{year}{2018}.
\bibitem[{Chen et~al.(2020)Chen, Zhang, and Peng}]{chen2020metier}
\bibinfo{author}{L.~Chen}, \bibinfo{author}{Y.~Zhang},
  \bibinfo{author}{L.~Peng},
\newblock \bibinfo{title}{Metier: A deep multi-task learning based activity and
  user recognition model using wearable sensors},
\newblock \bibinfo{journal}{Proceedings of the ACM on Interactive, Mobile,
  Wearable and Ubiquitous Technologies} \bibinfo{volume}{4}
  (\bibinfo{year}{2020}) \bibinfo{pages}{1--18}.
\bibitem[{Liu et~al.(2020)Liu, Yao, Li, Liu, Wang, Shao, and
  Abdelzaher}]{liu2020globalfusion}
\bibinfo{author}{S.~Liu}, \bibinfo{author}{S.~Yao}, \bibinfo{author}{J.~Li},
  \bibinfo{author}{D.~Liu}, \bibinfo{author}{T.~Wang},
  \bibinfo{author}{H.~Shao}, \bibinfo{author}{T.~Abdelzaher},
\newblock \bibinfo{title}{Giobalfusion: A global attentional deep learning
  framework for multisensor information fusion},
\newblock \bibinfo{journal}{Proceedings of the ACM on Interactive, Mobile,
  Wearable and Ubiquitous Technologies} \bibinfo{volume}{4}
  (\bibinfo{year}{2020}) \bibinfo{pages}{1--27}.
\bibitem[{Li et~al.(2021)Li, Cui, Wang, Zhang, Chen, and Wu}]{li2021two}
\bibinfo{author}{B.~Li}, \bibinfo{author}{W.~Cui}, \bibinfo{author}{W.~Wang},
  \bibinfo{author}{L.~Zhang}, \bibinfo{author}{Z.~Chen},
  \bibinfo{author}{M.~Wu},
\newblock \bibinfo{title}{Two-stream convolution augmented transformer for
  human activity recognition},
\newblock in: \bibinfo{booktitle}{Proceedings of the AAAI Conference on
  Artificial Intelligence}, volume~\bibinfo{volume}{35}, \bibinfo{year}{2021},
  pp. \bibinfo{pages}{286--293}.
\bibitem[{Islam and Iqbal(2022)}]{islam2022mumu}
\bibinfo{author}{M.~M. Islam}, \bibinfo{author}{T.~Iqbal},
\newblock \bibinfo{title}{Mumu: Cooperative multitask learning-based guided
  multimodal fusion,”},
\newblock \bibinfo{organization}{AAAI}, \bibinfo{year}{2022}.
\bibitem[{Suh et~al.(2023)Suh, Rey, and Lukowicz}]{suh2023tasked}
\bibinfo{author}{S.~Suh}, \bibinfo{author}{V.~F. Rey},
  \bibinfo{author}{P.~Lukowicz},
\newblock \bibinfo{title}{Tasked: Transformer-based adversarial learning for
  human activity recognition using wearable sensors via self-knowledge
  distillation},
\newblock \bibinfo{journal}{Knowledge-Based Systems} \bibinfo{volume}{260}
  (\bibinfo{year}{2023}) \bibinfo{pages}{110143}.
\bibitem[{Chen et~al.(2018)Chen, Zhang, Cao, and Guo}]{chen2018distilling}
\bibinfo{author}{Z.~Chen}, \bibinfo{author}{L.~Zhang},
  \bibinfo{author}{Z.~Cao}, \bibinfo{author}{J.~Guo},
\newblock \bibinfo{title}{Distilling the knowledge from handcrafted features
  for human activity recognition},
\newblock \bibinfo{journal}{IEEE Transactions on Industrial Informatics}
  \bibinfo{volume}{14} (\bibinfo{year}{2018}) \bibinfo{pages}{4334--4342}.
\bibitem[{Thoker and Gall(2019)}]{thoker2019cross}
\bibinfo{author}{F.~M. Thoker}, \bibinfo{author}{J.~Gall},
\newblock \bibinfo{title}{Cross-modal knowledge distillation for action
  recognition},
\newblock in: \bibinfo{booktitle}{2019 IEEE International Conference on Image
  Processing (ICIP)}, \bibinfo{organization}{IEEE}, \bibinfo{year}{2019}, pp.
  \bibinfo{pages}{6--10}.
\bibitem[{Gao et~al.(2020)Gao, Oh, Grauman, and Torresani}]{gao2020listen}
\bibinfo{author}{R.~Gao}, \bibinfo{author}{T.-H. Oh},
  \bibinfo{author}{K.~Grauman}, \bibinfo{author}{L.~Torresani},
\newblock \bibinfo{title}{Listen to look: Action recognition by previewing
  audio},
\newblock in: \bibinfo{booktitle}{Proceedings of the IEEE/CVF Conference on
  Computer Vision and Pattern Recognition}, \bibinfo{year}{2020}, pp.
  \bibinfo{pages}{10457--10467}.
\bibitem[{Liu et~al.(2021)Liu, Wang, Li, and Lin}]{liu2021semantics}
\bibinfo{author}{Y.~Liu}, \bibinfo{author}{K.~Wang}, \bibinfo{author}{G.~Li},
  \bibinfo{author}{L.~Lin},
\newblock \bibinfo{title}{Semantics-aware adaptive knowledge distillation for
  sensor-to-vision action recognition},
\newblock \bibinfo{journal}{IEEE Transactions on Image Processing}
  \bibinfo{volume}{30} (\bibinfo{year}{2021}) \bibinfo{pages}{5573--5588}.
\bibitem[{Ni et~al.(2022{\natexlab{a}})Ni, Ngu, and Yan}]{ni2022progressive}
\bibinfo{author}{J.~Ni}, \bibinfo{author}{A.~H. Ngu}, \bibinfo{author}{Y.~Yan},
\newblock \bibinfo{title}{Progressive cross-modal knowledge distillation for
  human action recognition},
\newblock in: \bibinfo{booktitle}{Proceedings of the 30th ACM International
  Conference on Multimedia}, \bibinfo{year}{2022}{\natexlab{a}}, pp.
  \bibinfo{pages}{5903--5912}.
\bibitem[{Ni et~al.(2022{\natexlab{b}})Ni, Sarbajna, Liu, Ngu, and
  Yan}]{ni2022cross}
\bibinfo{author}{J.~Ni}, \bibinfo{author}{R.~Sarbajna},
  \bibinfo{author}{Y.~Liu}, \bibinfo{author}{A.~H. Ngu},
  \bibinfo{author}{Y.~Yan},
\newblock \bibinfo{title}{Cross-modal knowledge distillation for
  vision-to-sensor action recognition},
\newblock in: \bibinfo{booktitle}{ICASSP 2022-2022 IEEE International
  Conference on Acoustics, Speech and Signal Processing (ICASSP)},
  \bibinfo{organization}{IEEE}, \bibinfo{year}{2022}{\natexlab{b}}, pp.
  \bibinfo{pages}{4448--4452}.
\bibitem[{Nagrani et~al.(2021)Nagrani, Yang, Arnab, Jansen, Schmid, and
  Sun}]{nagrani2021attention}
\bibinfo{author}{A.~Nagrani}, \bibinfo{author}{S.~Yang},
  \bibinfo{author}{A.~Arnab}, \bibinfo{author}{A.~Jansen},
  \bibinfo{author}{C.~Schmid}, \bibinfo{author}{C.~Sun},
\newblock \bibinfo{title}{Attention bottlenecks for multimodal fusion},
\newblock \bibinfo{journal}{Advances in Neural Information Processing Systems}
  \bibinfo{volume}{34} (\bibinfo{year}{2021}) \bibinfo{pages}{14200--14213}.
\bibitem[{Hinton et~al.(2014)Hinton, Vinyals, and Dean}]{hinton2015distilling}
\bibinfo{author}{G.~Hinton}, \bibinfo{author}{O.~Vinyals},
  \bibinfo{author}{J.~Dean},
\newblock \bibinfo{title}{Distilling the knowledge in a neural network},
\newblock \bibinfo{journal}{NIPS Deep Learning and Representation Learning
  Workshop}  (\bibinfo{year}{2014}).
\bibitem[{Islam et~al.(2022)Islam, Zang, Tomar, Didolkar, Islam, Arnob, Iqbal,
  Li, Goyal, Heess et~al.}]{islam2022representation}
\bibinfo{author}{R.~Islam}, \bibinfo{author}{H.~Zang},
  \bibinfo{author}{M.~Tomar}, \bibinfo{author}{A.~Didolkar},
  \bibinfo{author}{M.~M. Islam}, \bibinfo{author}{S.~Y. Arnob},
  \bibinfo{author}{T.~Iqbal}, \bibinfo{author}{X.~Li},
  \bibinfo{author}{A.~Goyal}, \bibinfo{author}{N.~Heess}, et~al.,
\newblock \bibinfo{title}{Representation learning in deep rl via discrete
  information bottleneck},
\newblock \bibinfo{journal}{arXiv preprint arXiv:2212.13835}
  (\bibinfo{year}{2022}).

\end{thebibliography}





\end{document}